\documentstyle[sprocl]{article}
\input epsf

\def\spose#1{\hbox to 0pt{#1\hss}}
\def\lsim{\mathrel{\spose{\lower 3pt\hbox{$\mathchar"218$}}
 \raise 2.0pt\hbox{$\mathchar"13C$}}}
\def\gsim{\mathrel{\spose{\lower 3pt\hbox{$\mathchar"218$}}
 \raise 2.0pt\hbox{$\mathchar"13E$}}}

\begin{document}

\begin{titlepage}

\begin{flushright}
CERN-TH/97-99\\
HD-THEP-97-23\\
hep-ph/9705292
\end{flushright}

\vspace{1cm}

\begin{center}
\Large\bf Non-Leptonic Weak Decays of B Mesons
\end{center}

\vspace{1cm}

\begin{center}
Matthias Neubert\\
{\sl Theory Division, CERN, CH-1211 Geneva 23, Switzerland}\\[0.2cm]
and\\[0.2cm]
Berthold Stech\\
{\sl Institut f\"ur Theoretische Physik der Universit\"at
Heidelberg\\
Philosophenweg 16, D-69120 Heidelberg, Germany}
\end{center}

\vspace{1cm}

\begin{abstract}
We present a detailed study of non-leptonic two-body decays of $B$
mesons based on a generalized factorization hypothesis. We discuss the
structure of non-factorizable corrections and present arguments in
favour of a simple phenomenological description of their effects. To
evaluate the relevant transition form factors in the factorized decay
amplitudes, we use information extracted from semileptonic decays and
incorporate constraints imposed by heavy-quark symmetry. We discuss
tests of the factorization hypothesis and show how unknown decay
constants may be determined from non-leptonic decays. In particular, we
find $f_{D_s}=(234\pm 25)$~MeV and $f_{D_s^*}=(271\pm 33)$~MeV.
\end{abstract}

\vspace{1cm}

\begin{center}
To appear in the Second Edition of\\
Heavy Flavours, edited by A.J. Buras and M. Lindner\\
(World Scientific, Singapore)
\end{center}

\vspace{1.5cm}

\noindent
CERN-TH/97-99\\
May 1997
\vfil

\end{titlepage}

\thispagestyle{empty}
\vbox{}
\newpage

\setcounter{page}{1}


\title{NON-LEPTONIC WEAK DECAYS OF B MESONS}

\author{MATTHIAS NEUBERT}

\address{Theory Division, CERN, CH-1211 Geneva 23, Switzerland}

\author{BERTHOLD STECH}

\address{Institut f\"ur Theoretische Physik der Universit\"at
Heidelberg\\
Philosophenweg 16, D-69120 Heidelberg, Germany}

\maketitle\abstracts{
We present a detailed study of non-leptonic two-body decays of $B$
mesons based on a generalized factorization hypothesis. We discuss the
structure of non-factorizable corrections and present arguments in
favour of a simple phenomenological description of their effects. To
evaluate the relevant transition form factors in the factorized decay
amplitudes, we use information extracted from semileptonic decays and
incorporate constraints imposed by heavy-quark symmetry. We discuss
tests of the factorization hypothesis and show how unknown decay
constants may be determined from non-leptonic decays. In particular, we
find $f_{D_s}=(234\pm 25)$~MeV and $f_{D_s^*}=(271\pm 33)$~MeV.}

\section{Introduction}

The weak decays of hadrons containing a heavy quark offer the most
direct way to determine the weak mixing angles of the
Cabibbo-Kobayashi-Maskawa (CKM) matrix and to explore the physics of
CP violation. At the same time, they are of great help in studying
strong-interaction physics related with the confinement of quarks and
gluons into hadrons. Indeed, both tasks complement each other: an
understanding of the connection between quark and hadron properties is
a necessary prerequisite for a precise determination of the CKM matrix
and CP-violating phases.

The simplest processes are those involving a minimum number of
hadrons, i.e.\ a single hadron in the final state of a semileptonic
decay, or two hadrons in the final state of a non-leptonic decay. In
recent years, much progress has been achieved in understanding these
processes. Simple bound-state models are able to describe, in a
semiquantitative way, the current matrix elements occurring in
semileptonic decay amplitudes \cite{ba85}$^-$\cite{Faus}. A
factorization prescription for reducing the hadronic matrix elements of
four-quark operators to products of current matrix elements shed light
onto the dynamics of non-leptonic processes \cite{bs85,bsw87}, where
even drastic effects had been lacking an explanation before.

More recently, the discovery of heavy-quark symmetry
\cite{Shu1}$^-$\cite{review} and the establishment of the heavy-quark
effective theory \cite{EiFe}$^-$\cite{vcb} have provided a solid
theoretical framework to calculate exclusive semileptonic transitions
between two hadrons containing heavy quarks, such as the decays $\bar
B\to D^{(*)}\ell\,\bar\nu$. Moreover, the heavy-quark expansion has
been applied to the calculation of inclusive semileptonic, non-leptonic
and rare decay rates \cite{Chay}$^-$\cite{Fermi} and of the lifetimes
of charm and bottom hadrons \cite{liferef}$^-$\cite{Beijing}. These
developments are discussed in detail in the article of one of us
(M.~Neubert) in this volume. In particular, they lead to a precise
determination of the CKM matrix element governing the strength of $b\to
c$ transitions:
\begin{equation}
   |V_{cb}| = 0.039\pm 0.002 \,.
\label{Vcbval}
\end{equation}

In this article, we shall discuss non-leptonic two-body decays of $B$
mesons. The dynamics of non-leptonic decays, in which only hadrons
appear in the final state, is strongly influenced by the confining
colour forces among the quarks. Whereas in semileptonic transitions the
long-distance QCD effects are described by few hadronic form factors
parametrizing the hadronic matrix elements of quark currents,
non-leptonic processes are complicated by the phenomenon of quark
rearrangement due to the exchange of soft and hard gluons. The
theoretical description involves matrix elements of local four-quark
operators, which are much harder to deal with than current operators.
These strong-interaction effects prevented for a long time a coherent
understanding of non-leptonic decays. The $\Delta I=\frac{1}{2}$ rule
in strange particle decays is a prominent example. Although this
selection rule had been known for almost four decades, only recently
successful theoretical approaches have been developed to explain it in
a semiquantitative way \cite{DeltaI}$^-$\cite{diquarks}. The strong
colour force between two quarks in a colour-antitriplet state has been
identified as the dominant mechanism responsible for the dramatic
enhancement of $\Delta I=\frac{1}{2}$ processes.

The discovery of the heavy charm and bottom quarks opened up the
possibility to study a great variety of new decay channels. By now,
there is an impressive amount of experimental data available on many
exclusive and inclusive decay modes. In many respects, non-leptonic
decays of heavy mesons are an ideal instrument for exploring the most
interesting aspect of QCD, i.e.\ its non-perturbative, long-range
character. Since the initial state consists of an isolated heavy
particle and the weak transition operator exhibits a well-known and
simple structure, a detailed analysis of decays into particles with
different spin and flavour quantum numbers provides valuable
information about the nature of the long-range forces influencing
these processes, the same forces that determine the internal
structure of all hadrons.

In energetic two-body transitions, hadronization of the decay
products does not occur until they have traveled some distance away
from each other~\cite{Bj89}. The reason is that once the quarks have
grouped into colour-singlet pairs, soft gluons are ineffective in
rearranging them. The decay amplitudes are then expected to factorize
into products of hadronic matrix elements of colour-singlet quark
currents. The factorization approximation has been applied to many
two-body decays of $B$ and $D$
mesons~\cite{bsw87,Nuss,Desh}$^-$\cite{Eber}. It relates the
complicated non-leptonic decay amplitudes to products of meson decay
constants and hadronic matrix elements of current operators, which are
similar to those encountered in semileptonic decays. The decay
constants are fundamental hadronic parameters providing a measure of
the strength of the quark-antiquark attraction inside a hadronic
state. As some of them are not directly accessible in leptonic or
electromagnetic processes, their extraction from non-leptonic
transitions may provide important information.

In Section~\ref{sec:Heff}, we discuss the effective Hamiltonian
relevant for decays of the bottom quark. In Section~\ref{sec:fact}, we
provide a prescription for the calculation of non-leptonic decay
amplitudes in the factorization approximation. We discuss the problems
connected with factorization, such as the choice of a suitable
factorization scale and its possible dependence on the energy released
in a decay process. In Section~\ref{sec:nonfac}, we then discuss in
more detail the structure of non-factorizable corrections. Using the
$1/N_c$ expansion, we argue that in energetic two-body decays of $B$
mesons a generalized factorization prescription holds, which involves
two parameters $a_1^{\rm eff}\approx c_1+\zeta c_2$ and $a_2^{\rm
eff}\approx c_2+\zeta c_1$ depending on a hadronic parameter
$\zeta=O(1/N_c)$. We estimate that the process dependence of $\zeta$ is
very mild, suppressed as $\Delta E/m_b$, where $\Delta E$ is the
difference in the energy release in different two-body decay channels.
Some basic aspects of final-state interactions are briefly covered in
Section~\ref{sec:FSI}. The remainder of this article is devoted to a
phenomenological description of non-leptonic two-body decays of $B$
mesons. In the factorization approximation, the ingredients for such an
analysis are meson decay constants and transition form factors. In
Section~\ref{sec:decay}, we collect the current experimental
information on meson decay constants. In Section~\ref{sec:formfa}, we
describe two simple models that provide a global description of
heavy-to-heavy and heavy-to-light weak decay form factors, embedding
the known constraints imposed by heavy-quark
symmetry~\cite{Isgu,review,Rieck}. With only a few parameters, these
models reproduce within errors the known properties of decay form
factors and predict those form factors which are yet unknown. Our
approach is meant to provide global predictions for a large set of
non-leptonic decay amplitudes. We do not attempt here a more
field-theoretical calculation of weak amplitudes, nor do we perform a
dedicated investigation of individual decay channels.\footnote{Attempts
of a more rigorous calculation of some particular decay modes are
discussed in the article by R.~R\"uckl in this volume.}
In Sections~\ref{sec:comparison} and \ref{sec:tests}, we compare our
predictions with the experimental data on two-body non-leptonic decays
of $B$ mesons, present tests of the factorization assumption, and
extract the decay constants of $D_s$ and $D_s^*$ mesons. In
Section~\ref{sec:baryons}, we briefly address the interesting
possibility of $B$-meson decays into baryons. A summary and conclusions
are given in Section~\ref{sec:sum}.

\section{Effective Hamiltonian}
\label{sec:Heff}

At tree level, non-leptonic weak decays are described in the Standard
Model by a single $W$-exchange diagram. Strong interactions affect this
simple picture in a two-fold way. Hard-gluon corrections can be
accounted for by perturbative methods and renormalization-group
techniques. They give rise to new effective weak vertices.
Long-distance confinement forces are responsible for the binding of
quarks inside the asymptotic hadron states. The basic tool in the
calculation of non-leptonic amplitudes is to separate the two regimes
by means of the operator product expansion (OPE)~\cite{wi69},
incorporating all long-range QCD effects in the hadronic matrix
elements of local four-quark operators. This treatment appears well
justified due to the vastly different time and energy scales involved
in the weak decay and in the subsequent formation of the final hadrons.

Integrating out the heavy $W$-boson and top-quark fields, one derives
the effective Hamiltonian for $b\to c,u$
transitions~\cite{gi79,Sh77}:
\begin{eqnarray}\label{Heff}
   H_{\rm eff} &=& {G_F\over\sqrt{2}}\,\bigg\{ V_{cb}\,\left[
    c_1(\mu)\,Q_1^{cb} + c_2(\mu)\,Q_2^{cb} \right] \nonumber\\
   &&\mbox{}+ V_{ub}\,\left[
    c_1(\mu)\,Q_1^{ub} + c_2(\mu)\,Q_2^{ub} \right]
    + \mbox{h.c.} \bigg\} \nonumber\\
   &&\mbox{}+ \hbox{penguin operators} .
\end{eqnarray}
It consists of products of local four-quark operators renormalized at
the scale $\mu$ and scale-dependent Wilson coefficients $c_i(\mu)$.
$V_{cb}$ and $V_{ub}$ are elements of the quark mixing matrix. The
operators $Q_1$ and $Q_2$, written as products of colour-singlet
currents, are given by
\begin{eqnarray}\label{Obc}
   Q_1^{cb} &=& \left[ (\bar d'u)_{V-A} + (\bar s'c)_{V-A} \right]
   (\bar c b)_{V-A} \,, \nonumber\\[0.1cm]
   Q_2^{cb} &=& (\bar c u)_{V-A}\,(\bar d'b)_{V-A}
    + (\bar c c)_{V-A}\,(\bar s'b)_{V-A} \,, \nonumber\\[0.15cm]
   Q_1^{ub} &=& \left[ (\bar d'u)_{V-A} + (\bar s'c)_{V-A} \right]
   (\bar u b)_{V-A} \,, \nonumber\\[0.1cm]
   Q_2^{ub} &=& (\bar u u)_{V-A}\,(\bar d'b)_{V-A}
    + (\bar u c)_{V-A}\,(\bar s'b)_{V-A} \,,
\end{eqnarray}
where $d'$ and $s'$ denote weak eigenstates of the down and strange
quarks, respectively, and $(\bar c b)_{V-A}=\bar c\,\gamma_\mu
(1-\gamma_5) b$ etc. Without strong-interaction effects we would have
$c_1=1$ and $c_2=0$. This simple result is modified, however, by gluon
exchange: the original weak vertices get renormalized, and new types of
interactions (such as the operators $Q_2$) are induced. Not explicitly
shown in (\ref{Heff}) are the so-called penguin operators~\cite{Sh77}.
Since their Wilson coefficients are very small, the corresponding
contributions to weak decay amplitudes only become relevant in rare
decays, where the tree-level contribution is either strongly
CKM-suppressed, as in $\bar B\to\bar K^{(*)}\pi$, or where matrix
elements of the $Q_1$ and $Q_2$ operators do not contribute at all, as
in $\bar B\to\bar K^*\gamma$ and $\bar B^0\to\bar K^0\phi$.

\begin{figure}
\epsfysize=3cm
\centerline{\epsffile{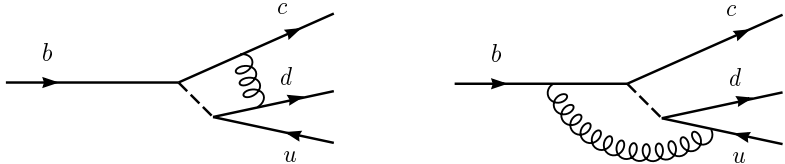}}
\caption{\label{Fig:gluons}
Hard-gluon corrections giving rise to the Wilson coefficients
$c_1(\mu)$ and $c_2(\mu)$ in the effective weak Hamiltonian.}
\end{figure}

The Wilson coefficients $c_i(\mu)$ take into account the short-distance
corrections arising from the exchange of gluons with virtualities
between $m_W$ and some hadronic scale $\mu$, chosen large enough for
perturbation theory to be applicable. The coefficients $c_1(\mu)$ and
$c_2(\mu)$, for example, arise from the hard-gluon exchanges shown in
Figure~\ref{Fig:gluons}. The effects of soft gluons (with virtualities
below the scale $\mu$) remain in the hadronic matrix elements of the
local four-quark operators $Q_i$. In the course of the evolution from
$m_W$ down to the scale $\mu$ there arise large logarithms of the type
$[\alpha_s\ln(m_W/\mu)]^n$, which must be summed to all orders in
perturbation theory. This is achieved by means of the
renormalization-group equation (RGE). The combinations
$c_{\pm}(\mu)=c_1(\mu)\pm c_2(\mu)$ of the Wilson coefficients have a
multiplicative evolution under change of the renormalization scale.
They can be obtained from the solution of the RGE
\begin{equation}\label{RGE}
   \left( \mu\,{{\rm d}\over{\rm d}\mu} - \Gamma_\pm \right)
   c_\pm(\mu) = 0 \,,
\end{equation}
with the initial condition $c_\pm(m_W)=1$, corresponding to $c_1(m_W)
=1$ and $c_2(m_W)=0$. The quantities $\Gamma_\pm$ in (\ref{RGE}) are
the anomalous dimensions of the operators $(Q_1\pm Q_2)$. At one-loop
order, they are given by~\cite{Al74,GaLe}
\begin{equation}\label{gplgmi}
   \Gamma_\pm = \gamma_\pm\,{\alpha_s\over 4\pi} \,;\qquad
   \gamma_\pm = 6 \left( \pm 1 - {1\over N_c} \right) \,,
\end{equation}
where $N_c=3$ is the number of colours. To leading logarithmic order
(LO), the solution of the RGE is
\begin{equation}\label{cpm}
   c_{\pm}(\mu) = \left( {\alpha_s(m_W)\over\alpha_s(\mu)}
   \right)^{\gamma_{\pm}/2\beta_0} \,;\qquad
   \beta_0 = {11\over 3}\,N_c - {2\over 3}\,n_f \,,
\end{equation}
where $\beta_0$ is the first coefficient of the $\beta$ function, and
$n_f$ the number of active flavours (in the region between $m_W$ and
$\mu$). The physical origin of the enhancement of $c_-(\mu)$ can be
traced back to the attractive force between two quarks in the
colour-antitriplet channel of the scattering process $b+u\to c+d$.
Similarly, in the colour-sextet channel the force is repulsive, leading
to $c_+(\mu )<1$.

The leading logarithmic approximation can be improved by including the
next-to-leading (NLO) corrections of order
$\alpha_s[\alpha_s\ln(m_W/\mu)]^n$. The result is
\begin{equation}
   c_{\pm}(\mu) = \left( {\alpha_s(m_W)\over\alpha_s(\mu)}
   \right)^{\gamma_{\pm}/2\beta_0}\,\left\{ 1 +
   R_\pm\,{\alpha_s(\mu)-\alpha_s(m_W)\over 4\pi} \right\} \,,
\end{equation}
where the coefficients $R_\pm$ are given by~\cite{ACMP,BuWe}
($\beta_1$ is the two-loop coefficient of the $\beta$ function)
\begin{equation}
   R_\pm = {N_c\mp 1\over 2 N_c}\,\left\{ \pm {6\beta_1\over\beta_0^2}
   + {1\over 2\beta_0}\,\bigg( 21\mp {57\over N_c}\pm {19\over 3}\,
   N_c\mp {4\over 3}\,n_f \bigg) \mp 11 \right\} \,.
\end{equation}
In Table~\ref{tab:c1c2}, we show the values of $c_1(m_b)$ and
$c_2(m_b)$ obtained at leading and next-to-leading order. The evolution
of the running coupling constant is done at two-loop order, using the
normalization $\alpha_s(m_Z)=0.118\pm 0.003$. For comparison, we show
in Table~\ref{tab:charm} the values obtained at a lower scale, which
may be relevant to charm decays. The negative sign of $c_2(\mu)$ has
important consequences for decays in which there is significant
interference between $c_1$ and $c_2$ amplitudes.

\begin{table}
\caption{\label{tab:c1c2}
Values of the Wilson coefficients at the scale $m_b=4.8$~GeV, both at
leading (LO) and next-to-leading (NLO) order}
\vspace{0.4cm}
\begin{center}
\begin{tabular}{|c|cc|cc|}\hline
\rule[-0.2cm]{0cm}{0.6cm}
$\alpha_s(m_Z)$ & $c_1^{\rm LO}(m_b)$ & $c_2^{\rm LO}(m_b)$ &
 $c_1^{\rm NLO}(m_b)$ & $c_2^{\rm NLO}(m_b)$ \\
\hline
0.115 & 1.102 & $-0.239$ & 1.124 &
 \rule[-0.2cm]{0cm}{0.6cm} $-0.273$ \rule[-0.2cm]{0cm}{0.6cm} \\[0.1cm]
0.118 & 1.108 & $-0.249$ & 1.132 & $-0.286$ \\[0.1cm]
0.121 & 1.113 & $-0.260$ & 1.140 & $-0.301$ \\[0.1cm]
\hline
\end{tabular}
\end{center}
\vspace{0.5cm}
\caption{\label{tab:charm}
Values of the Wilson coefficients at the scale $m_c=1.4$~GeV}
\vspace{0.4cm}
\begin{center}
\begin{tabular}{|c|cc|cc|}\hline
\rule[-0.2cm]{0cm}{0.6cm}
$\alpha_s(m_Z)$ & $c_1^{\rm LO}(m_c)$ & $c_2^{\rm LO}(m_c)$ &
 $c_1^{\rm NLO}(m_c)$ & $c_2^{\rm NLO}(m_c)$ \\
\hline
0.115 & 1.240 & $-0.476$ & 1.313 &
 \rule[-0.2cm]{0cm}{0.6cm} $-0.576$ \rule[-0.2cm]{0cm}{0.6cm} \\[0.1cm]
0.118 & 1.263 & $-0.513$ & 1.351 & $-0.631$ \\[0.1cm]
0.121 & 1.292 & $-0.556$ & 1.397 & $-0.696$ \\[0.1cm]
\hline
\end{tabular}
\end{center}
\end{table}

In a typical hadronic decay process such as $\bar B^0\to D^+\varrho^-$,
there are several mass scales involved: the hadron masses, the quark
masses, the energy release, etc. Thus, there is an uncertainty in the
choice of the ``characteristic scale'' of a process. In principle this
is not a problem, since the products of the Wilson coefficients with
the hadronic matrix elements of the local four-quark operators are
scale independent. In practice, however, one often employs simple model
estimates of the matrix elements, which usually do not yield an
explicit scale dependence that could compensate for that of the Wilson
coefficients. Instead, these model calculations are assumed to be valid
on a particular scale. A related technical problem is that whereas the
operator evolution from $m_W$ down to $m_b$ can be calculated in a
straightforward way, a more complicated scaling behaviour is expected
below the mass of the $b$ quark, where all the other mass scales start
to become relevant. There have been attempts to account for the scaling
in the region $\mu<m_b$ by summing logarithms of the type
$[\alpha_s\ln(m_b/\mu)]^n$ and $[\alpha_s\ln(E/\mu)]^n$, with
$E=O(m_b)$ being the large energy of a light particle produced in a
two-body decay of a $B$ meson~\cite{G/W,E/mu}. However, at present the
treatment of such corrections is still associated with large
uncertainties. We shall thus stay with the conventional choice
$\mu=O(m_b)$ adopted in most previous work on non-leptonic $B$ decays.

\section{Factorization}
\label{sec:fact}

In weak interactions, a meson (or meson resonance) can be directly
generated by a quark current carrying the appropriate parity and
flavour quantum numbers. The corresponding contribution to a decay
amplitude factorizes into the product of two current matrix
elements~\cite{fact1,fact2}. As an example, consider the transition
$\bar B^0\to D^+\,\pi^-$. The factorizable part of the amplitude is
given by
\begin{equation}\label{Afact}
   A_{\rm fact} = - {G_F\over\sqrt{2}}\,V_{cb}\,V_{ud}^*\,a_1\,
   \langle\pi^-|\,(\bar d u)_A\,|\,0\,\rangle\,
   \langle D^+|\,(\bar c b)_V\,|\bar B^0\rangle \,.
\end{equation}
The coefficient $a_1$ will be discussed below. The $\bar B^0\to D^+$
transition matrix element is of the same type as that encountered in
the semileptonic decay $\bar B^0\to D^+\ell^-\nu$. It can be determined
using data on semileptonic decays together with theoretical arguments
based on heavy-quark symmetry~\cite{review}. The amplitude for creating
a pion from the vacuum via the axial current is parametrized by the
decay constant $f_\pi$ and is proportional to the momentum of the pion:
\begin{equation}\label{fpi}
   \langle\pi^-(p)|\,\bar d\,\gamma_\mu\gamma_5 u\,|\,0\,\rangle
   = i f_\pi p_{\mu} \,.
\end{equation}
Thus it seems natural to assume that the amplitude for energetic weak
decays, in which the directly generated meson carries a large
momentum, is dominated by its factorizable part. This assumption can
be substantiated by a more detailed analyses of the kinematic
situation in the above decay process: a fast moving $(\bar u d)$ pair
created in a point-like interaction, with both quarks leaving the
interaction region in the same direction and with a velocity close to
the speed of light, will hadronize only after a time given by its
$\gamma$ factor times a typical hadronization time $\tau_{\rm
had}\sim 1$~fm/c. In the above example, this means that hadronization
occurs about 20~fm away from the remaining quarks. Inside the
interaction region, the $(\bar u d)$ pair behaves like a colourless
and almost point-like particle. It does little interact with the
remaining quarks. Because of this intuitive ``colour transparency
argument''~\cite{Bj89}, one expects that the factorizable part
(\ref{Afact}) of the decay amplitude does indeed give the dominant
contribution to the full amplitude. A more formal investigation of
this situation, using an effective theory for heavy quarks and
fast-moving light quarks, has been presented in Ref.~72.

In the following applications, we shall adopt the factorization
ansatz also for processes in which the $\gamma$ factors of the
outgoing particles are not necessarily large. Examples of such decays
are $\bar B\to\bar K\,J/\psi$ and $\bar B\to D\,\bar D_s$. In these
cases, the kinematic argument given above does no longer apply.
Nevertheless, in a two-body process, the concentration of energy into
colour-singlet states together with the fact that (axial) vector
current matrix elements increase with the particle momentum favours
the direct current-induced production of mesons over more complicated
production mechanisms. Only comparison with experiment can tell whether
factorization is a useful concept also for those processes. We will
analyse the structure of non-factorizable corrections in
Section~\ref{sec:nonfac}, and discuss some tests of the factorization
approximation in Section~\ref{sec:tests}.

By factorizing matrix elements of the four-quark operators contained in
the effective Hamiltonian (\ref{Heff}), one can distinguish three
classes of decays~\cite{bs85,bsw87}. The first class contains those
decays in which only a charged meson can be generated directly from a
colour-singlet current, as in $\bar B^0\to D^+\pi^-$. For these
processes, the relevant QCD coefficient is given by the combination
\begin{equation}\label{a1}
   a_1 = c_1(\mu_f) + \zeta\,c_2(\mu_f) \,,\qquad \mbox{(class~I)}
\end{equation}
where $\zeta=1/N_c$ ($N_c$ being the number of quark colours), and
$\mu_f=O(m_b)$ is the scale at which factorization is assumed to be
relevant. The term proportional to $c_2(\mu_f)$ arises from the Fierz
reordering of $Q_2$ operators and factorization of the product of
colour-singlet currents contained in it. At this point, the
colour-octet term resulting from the Fierz reordering is simply being
discarded. In order to remove this deficiency, one should treat $\zeta$
as a free parameter. This will be discussed in more detail in the next
section.

A second class of transitions consists of those decays in which the
meson generated directly from the current is neutral, like the
$J/\psi$ particle in the decay $\bar B \to\bar K\,J/\psi$. The
corresponding decay amplitude,
\begin{equation}
   A_{\rm fact} = {G_F\over\sqrt{2}}\,V_{cb}\,V_{cs}^*\,a_2\,
   \langle J/\psi|\,(\bar c c)_V\,|\,0\,\rangle\,
    \langle\bar K|\,(\bar s b)_V\,|\bar B\rangle \,,
\end{equation}
is proportional to the QCD coefficient
\begin{equation}\label{a2}
  a_2 = c_2(\mu_f) + \zeta\,c_1(\mu_f) \,.\qquad \mbox{(class~II)}
\end{equation}
Due to the different sign and magnitude of the Wilson coefficients
(see Table~\ref{tab:c1c2}), the combination $a_2$ is particularly
sensitive to the value of the factorization scale and to any
additional long-distance contributions (i.e., to the precise value of
$\zeta$). The QCD coefficient $a_1$, on the other hand, can be
estimated quite reliably.

The third class of transitions covers decays in which the $a_1$ and
$a_2$ amplitudes interfere, such as in $B^-\to D^0\pi^-$. Their final
state contains a charged as well as a neutral meson, both of which can
be generated from a current of one of the operators of the effective
Hamiltonian. The corresponding amplitudes involve a combination
\begin{equation}
  a_1 + x\,a_2 \,,\qquad \mbox{(class~III)}
\end{equation}
where $x=1$ in the formal limit of a flavour symmetry for
the final-state mesons, as it is realized in the strongly
CKM-suppressed decay $B^-\to\pi^0\pi^-$.

In principle, there is another type of factorizable contribution to
weak decay amplitudes, which is however significantly different from
the ones covered so far: the so-called ``weak annihilation
contribution''~\cite{wA1,wA2}, in which the decaying heavy meson is
annihilated by a current of one of the operators of the effective
Hamiltonian. For a charged meson, this contribution is proportional to
$a_1$, while it is proportional to $a_2$ for a neutral meson. In the
latter case, the weak annihilation is in fact the exchange of a $W$
boson between the two constituent quarks. In a weak annihilation
process, the second current in the four-quark operator produces all the
recoiling final-state particles out of the vacuum, which implies a
sizable form factor suppression. Therefore, annihilation amplitudes are
expected to be small, and it is commonly assumed that they may be
neglected for all except some rare processes. This approximation is not
an essential part of the factorization scheme, however. It would be
straightforward to include the annihilation contributions if the
relevant form factors at large time-like values of $q^2$ were known.

Let us come back to our first example of a class~I decay, collecting
all the different pieces of the factorizable contribution to the
decay amplitude. Inserting (\ref{fpi}) and a suitable form factor
decomposition of the hadronic matrix element (see Appendix) into
(\ref{Afact}), we obtain
\begin{equation}\label{Afact2}
   A_{\rm fact}(\bar B^0\to D^+\pi^-)
   = -i\,{G_F\over\sqrt{2}}\,V_{cb}\,V_{ud}^*\,a_1\,f_{\pi}\,
   (m_B^2-m_D^2)\,F_0^{B\to D}(m_\pi^2) \,.
\end{equation}
Note that it is the longitudinal form factor $F_0$ which enters this
expression, whereas (in the limit of vanishing lepton mass) the
semileptonic decay amplitude for $\bar B^0\to D^+\ell^-\nu$ involves
the transverse form factor $F_1$. Fortunately, heavy-quark symmetry
provides relations between all the $\bar B\to D^{(*)}$ transition form
factors~\cite{review,Rieck}. They allow us to relate $F_0$ to form
factors which are easily accessible experimentally.\footnote{In
addition, in the present case the smallness of the pion mass allows us
to use the kinematic constraint $F_0(0)=F_1(0)$ (see Appendix) to
relate $F_0$ to a measurable semileptonic form factor.}

\begin{figure}
\epsfxsize=7cm
\centerline{\epsffile{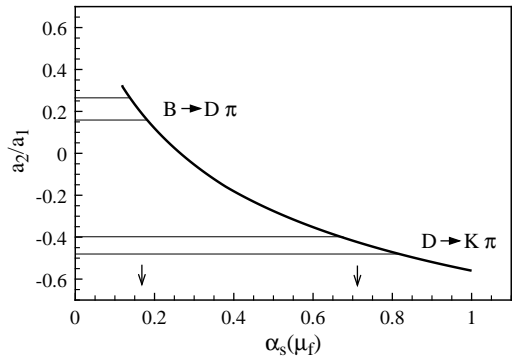}}
\caption{\label{Fig:a2a1}
The ratio $a_2/a_1$ as a function of the running coupling constant
evaluated at the factorization scale. The bands indicate the
phenomenological values of $a_2/a_1$ extracted from $\bar B\to D\pi$
and $D\to K\pi$ decays.}
\end{figure}

Note that the hadronic matrix elements of the vector and axial currents
resulting from the factorization of the matrix elements of four-quark
operators do not show any scale dependence that could compensate the
scale dependence of the Wilson coefficients. Strictly speaking,
therefore, factorization cannot be correct. What one may hope for is
that it provides a useful approximation if the Wilson coefficients (or
equivalently the QCD coefficients $a_1$ and $a_2$) are evaluated at a
suitable scale $\mu_f$, the factorization point. Because the factorized
hadronic matrix elements can only account for the interaction between
quarks remaining together in the same hadron, the Wilson coefficients
in effective Hamiltonian should contain those gluon effects which
redistribute the quarks. Thus, we have to evolve these coefficients
down to a scale where the gluons are no longer effective in changing
the particle momenta in a significant way. For very energetic decays,
the ``colour transparency argument'' shows that gluons with
virtualities much below the mass of the decaying particle are
ineffective in rearranging the final-state quarks~\cite{Bj89,E/mu}, and
$\mu_f=O(m_b)$ seems a reasonable choice for the factorization scale.
However, for processes in which the decay products carry only little
kinetic energy, gluons with virtualities well below $m_b$ can still
lead to a redistribution of the quarks before hadronization sets in.
This fact may be used as an argument in favour of a lower factorization
scale in such processes. On a qualitative level, the connection between
the factorization scale and the energy release in the final state can
be seen from Figure~\ref{Fig:a2a1}, where we show the ratio $a_2/a_1$
as a function of $\alpha_s(\mu_f)$. As we will see in
Section~\ref{sec:comparison}, the value preferred by $\bar B\to D\pi$
decays is positive and corresponds to a rather small coupling,
indicating $\mu_f=O(m_b)$. On the other hand, $D$ decays indicate a
negative value of $a_2/a_1$, corresponding to a significantly lower
value of the factorization scale. This is in accordance with the fact
that in these processes the energy released to the final-state
particles is only about 1~GeV. Moreover, it reflects that for
low-energetic processes hadronic weak decays exhibit a strong $\Delta
I=\frac 12$ enhancement, which is significant in $D$ decays and even
spectacular in $K$ decays. A strict $\Delta I=\frac 12$ selection rule
would correspond to $a_2/a_1=-1$. Because of these intuitive arguments,
we are led to abandon the ``naive'' factorization prescription adopted
in most previous work, where a fixed factorization scale $\mu_f=m_b$
was taken for all $B$ decays, in favour of a more flexible
factorization scheme, in which $\mu_f$ -- or equivalently $\zeta$ in
(\ref{a1}) and (\ref{a2}) -- is treated as a process-dependent
parameter.

We end this section with a remark on very low-energetic processes such
as $K$ decays. Factorization of hadronic matrix elements of four-quark
operators into two matrix elements of colour-singlet currents implies
that only those non-perturbative forces that act between quarks and
antiquarks are taken into account. The remaining interactions
including, in particular, the gluon exchange between two quarks or two
antiquarks are treated perturbatively. However, in the decays of
strange particles the long-distance attraction between two quarks in a
colour-antitriplet state gives the dominant contribution to $\Delta
I=\frac 12$ decay amplitudes. A detailed analysis of these ``diquark
effects'' can be found in Ref.~51.
As will be briefly
discussed in Section~\ref{sec:baryons}, we expect similar effects to be
important in $B$ decays into baryon-antibaryon pairs. For energetic $B$
decays into two mesons, on the other hand, the two quarks (or
antiquarks) do not end up in the same final-state hadron, and a
perturbative treatment of their mutual interaction seems justified.

\boldmath
\section{Corrections to Factorization and the 1$/N_c$ Expansion}
\unboldmath
\label{sec:nonfac}

Let us now illustrate in more detail the structure of the corrections
to the factorization approximation (for a similar discussion, see
Refs.~59, 60).
For a given class~I or class~II two-body decay
channel, the effective Hamiltonian (\ref{Heff}) can be rewritten using
Fierz identities in such a way that the quarks are paired according to
the flavour quantum numbers of the final-state hadrons.\footnote{In all
our applications, class~III amplitudes are related to combinations of
class~I and II amplitudes by isospin relations.}
This introduces products of two colour-singlet or two colour-octet
current operators. The hadronic matrix elements of the latter ones
are being neglected in the naive factorization approximation.

Consider again the example of the decay $\bar B^0\to D^+\pi^-$. In
this case, the appropriate form to write the effective Hamiltonian is
\begin{eqnarray}
   H_{\rm eff} = {G_F\over\sqrt{2}}\,V_{cb}\,V_{ud}^* &\bigg\{&
    \!\left( c_1(\mu) + {c_2(\mu)\over N_c} \right)\,
    (\bar d u)_{V-A}\,(\bar c b)_{V-A} \nonumber\\
   &&\mbox{}+ {c_2(\mu)\over 2}\,(\bar d t_a u)_{V-A}\,
    (\bar c t_a b)_{V-A} \bigg\} + \dots \,,
\end{eqnarray}
where $(\bar d\,t_a u)_{V-A}=\bar d\,\gamma_\mu(1-\gamma_5) t_a u$
etc., and we have shown explicitly the dependence on the number of
colours. For a given decay process, let us now define two hadronic
parameters as follows:
\begin{eqnarray}
   \varepsilon_1^{(BD,\pi)}(\mu)
   &\equiv& {\langle\pi^- D^+|\,(\bar d u)_{V-A}\,
    (\bar c b)_{V-A}\,|\bar B^0\rangle\over
    \langle\pi^-|\,(\bar d u)_{V-A}\,|\,0\,\rangle\,
    \langle D^+|\,(\bar c b)_{V-A}\,|\bar B^0\rangle} - 1 \,,
    \nonumber\\
   \varepsilon_8^{(BD,\pi)}(\mu)
   &\equiv& {\langle\pi^- D^+|\,(\bar d t_a u)_{V-A}\,
    (\bar c t_a b)_{V-A}\,|\bar B^0\rangle\over
    2 \langle\pi^-|\,(\bar d u)_{V-A}\,|\,0\,\rangle\,
    \langle D^+|\,(\bar c b)_{V-A}\,|\bar B^0\rangle} \,.
\end{eqnarray}
They parametrize the non-factorizable contributions to the hadronic
matrix elements and are process dependent. Here the subscript refers to
the colour structure of the operator, whereas the superscript indicates
the particles involved in the process. The final-state particle with
flavour quantum numbers of one of the currents is given last. This
superscript is written only when necessary to avoid confusion. In terms
of these new parameters the decay amplitude resumes the form given in
(\ref{Afact}), however, with $a_1$ replaced by the new coefficient
\begin{equation}\label{a1eff}
   a_1^{\rm eff} = \left( c_1(\mu) + {c_2(\mu)\over N_c} \right)
   \Big[ 1 + \varepsilon_1^{(BD,\pi)}(\mu) \Big]
   + c_2(\mu)\,\varepsilon_8^{(BD,\pi)}(\mu) \,.
\end{equation}
Similarly, in the class~II transition $\bar B^0\to D^0\pi^0$ we
encounter the effective coefficient
\begin{equation}\label{a2eff}
   a_2^{\rm eff} = \left( c_2(\mu) + {c_1(\mu)\over N_c} \right)
   \Big[ 1 + \varepsilon_1^{(B\pi,D)}(\mu) \Big]
   + c_1(\mu)\,\varepsilon_8^{(B\pi,D)}(\mu) \,.
\end{equation}

Since we have introduced these parameters without loss of generality,
the effective coefficients $a_i^{\rm eff}$ take into account all
contributions to the matrix elements and are thus $\mu$ independent.
In other words, the hadronic parameters $\varepsilon_i(\mu)$ restore
the correct $\mu$ dependence of the matrix elements, which is lost in
the naive factorization approximation. Using the RGE for the
coefficient functions, it is then straightforward to
show that the $\mu$ dependence of the hadronic parameters is, in the
leading logarithmic order, given by
\begin{eqnarray}
   1 + \varepsilon_1(\mu) &=& {1\over 2}\,\bigg[
    \bigg( 1 + {1\over N_c} \bigg) \Big[ 1 + \varepsilon_1(\mu_0)
    \Big] + \varepsilon_8(\mu_0) \bigg]\,\kappa_+ \nonumber\\
   &+& {1\over 2}\,\bigg[
    \bigg( 1 - {1\over N_c} \bigg) \Big[ 1 + \varepsilon_1(\mu_0)
    \Big] - \varepsilon_8(\mu_0) \bigg]\,\kappa_- \,, \nonumber\\
   \varepsilon_8(\mu) &=& {1\over 2}\,\bigg[
    \bigg( 1 - {1\over N_c} \bigg)\,\varepsilon_8(\mu_0)
    + \bigg( 1 - {1\over N_c^2} \bigg) \Big[
    1 + \varepsilon_1(\mu_0) \Big] \bigg]\,\kappa_+ \nonumber\\
   &+& {1\over 2}\,\bigg[
    \bigg( 1 + {1\over N_c} \bigg)\,\varepsilon_8(\mu_0)
    - \bigg( 1 - {1\over N_c^2} \bigg) \Big[
    1 + \varepsilon_1(\mu_0) \Big] \bigg]\,\kappa_- \,, \nonumber\\
\end{eqnarray}
where $\kappa_\pm=[\alpha_s(\mu)/\alpha_s(\mu_0)]^{\gamma_\pm/
2\beta_0}$, and $\mu_0$ is an arbitrary normalization point. It is
interesting to evaluate these expressions assuming that there exists a
scale $\mu_0=\mu_f$ where factorization holds, i.e.\ where
$\varepsilon_i(\mu_f)=0$. It then follows that
\begin{eqnarray}
   \varepsilon_1(\mu) &=& {1\over 2}\,\bigg( 1 + {1\over N_c}
    \bigg)\,\kappa_+ + {1\over 2}\,\bigg( 1 - {1\over N_c} \bigg)\,
    \kappa_- - 1 = O(1/N_c^2) \,, \nonumber\\
   \varepsilon_8(\mu) &=& {1\over 2}\,\bigg( 1 - {1\over N_c^2}
    \bigg) (\kappa_+ - \kappa_-) = O(1/N_c) \,,
\end{eqnarray}
and expanding in powers of $\alpha_s$ we obtain
\begin{equation}
   \varepsilon_1(\mu) = O(\alpha_s^2) \,, \qquad
   \varepsilon_8(\mu) = - {4\alpha_s\over 3\pi}\,\ln{\mu\over\mu_f}
   + O(\alpha_s^2) \,.
\label{mudepe}
\end{equation}
To deduce the dependence on the number of colours, we have used the
expressions for the anomalous dimension given in (\ref{gplgmi}). Note
that the large-$N_c$ counting rules, $\varepsilon_1=O(1/N_c^2)$ and
$\varepsilon_8=O(1/N_c)$, are independent of the assumption that
factorization holds at the scale $\mu_f$~\cite{Witt,BGR}. Hence, from
first principles of QCD, we expect that $|\varepsilon_1|\ll 1$, whereas
contributions from $\varepsilon_8$ can be more sizable.

Similar counting rules can be derived for the Wilson coefficients
$c_i(\mu)$; however, they can be obscured by the presence of the large
logarithm $L=\ln(m_W/\mu)$. In general, we have
\begin{equation}
   c_1(\mu) = 1 + O(L/N_c^2) \,, \qquad c_2(\mu) = O(L/N_c) \,.
\end{equation}
At the scale $\mu=m_b$, it is evident from Table~\ref{tab:c1c2} that
one can consistently treat $L=O(1)$, and thus $c_1=1+O(1/N_c^2)$ and
$c_2=O(1/N_c)$. For much lower scales, such as $\mu=m_c$, however,
Table~\ref{tab:charm} shows that it is appropriate to take
$L/N_c=O(1)$, and therefore $c_1=1+O(1/N_c)$ and $c_2=O(1)$.

Let us now discuss the relation of this general approach to the
conventional factorization scheme, focusing first on the case of $B$
decays. Naive factorization corresponds to setting
$\varepsilon_i(m_b)=0$, in which case $a_1$ and $a_2$ are universal,
process-independent coefficients, which are simply linear combinations
of the Wilson coefficients $c_1(m_b)$ and $c_2(m_b)$. In (\ref{a1eff})
and (\ref{a2eff}) we have shown how these relations are modified by the
presence of non-factorizable corrections. From the above discussion it
follows that non-factorizable corrections are expected to be small in
class~I transitions. In class~II decays, on the other hand, the
contribution proportional to $\varepsilon_8$ is enhanced by the large
value of the ratio $|c_1/c_2|$. Hence, as a general rule, we expect
sizable violations of the naive factorization approximation in class~II
decays only. Explicitly, evaluating the general expressions in the
large-$N_c$ limit, we find
\begin{eqnarray}
   a_1^{\rm eff} &=& c_1(\mu) + c_2(\mu) \left( \frac{1}{N_c}
    + \varepsilon_8^{(BD,\pi)}(\mu) \right) + O(1/N_c^2)
    = 1 + O(L/N_c^2) \,, \nonumber\\
   a_2^{\rm eff} &=& c_2(\mu) + c_1(\mu) \left( \frac{1}{N_c}
    + \varepsilon_8^{(B\pi,D)}(\mu) \right) + O(L/N_c^3) \,.
\end{eqnarray}
In $B$ decays, with $L=O(1)$ and neglecting terms of order $1/N_c^2$,
we can rewrite this as
\begin{equation}\label{a1a2appr}
   a_1^{\rm eff} \approx 1 \,,\qquad
   a_2^{\rm eff} \approx c_2(m_b) + \zeta\,c_1(m_b) \,,
\end{equation}
with
\begin{equation}
   \zeta \equiv {1\over N_c} + \varepsilon_8^{(B\pi,D)}(m_b) \,.
\end{equation}
Note that, with the same accuracy, the first relation in
(\ref{a1a2appr}) may be replaced by $a_1^{\rm eff}\approx
c_1(m_b)+\zeta\,c_2(m_b)$. It is important to stress at this point that
the naive choice $a_1=c_1+c_2/N_c$ and $a_2=c_2+c_1/N_c$ does not
correspond to any consistent limit of QCD~\cite{BGR}; in particular,
this is not a prediction of the $1/N_c$ expansion. Since the parameter
$\varepsilon_8$ is of order $1/N_c$, the two contributions to $\zeta$
are expected to be of similar magnitude, and hence $\zeta$ should be
considered as an unknown dynamical parameter. In other words, the
large-$N_c$ counting rules of QCD predict that related class~I and
class~II two-body decays, such as $\bar B^0\to D^+\pi^-$ and $\bar
B^0\to D^0\pi^0$, or $\bar B^0\to D^+\varrho^-$ and $\bar B^0\to
D^0\varrho^0$, can be described (up to corrections of order $1/N_c^2$)
using factorization with a single phenomenological parameter
$\zeta=O(1/N_c)$. Strictly speaking, however, this parameter will take
different values for different class~II decay channels. The picture
emerging from this discussion is equivalent to the concept of using a
process-dependent factorization scale $\mu_f$, discussed in the
previous section. This scale is defined such that
$\varepsilon_8(\mu_f)\equiv 0$. Using (\ref{mudepe}) one may then
calculate the corresponding value of $\varepsilon_8(m_b)$ and thus the
$\zeta$ parameter for each process.

We can go a step further and estimate the expected size of the process
dependence of the parameter $\zeta$. From Figure~\ref{Fig:a2a1}, we
know that for the particular case of $\bar B\to D\pi$ transitions
factorization holds at a high scale, i.e.\
$\varepsilon_8^{(B\pi,D)}(\mu_f)=0$ for $\mu_f\gsim m_b$ and thus
$\zeta\approx 1/3$. If we now assume that in other two-body decays of
$B$ mesons the factorization scale is lower because there is less
energy released to the final-state particles, we may use relation
(\ref{mudepe}) to obtain for the change in $\zeta$:
\begin{equation}
   \Delta\zeta = \Delta\varepsilon_8(m_b) \approx
   {4\alpha_s(m_b)\over 3\pi}\,\frac{\Delta E}{m_b}
   \approx 0.02\times\frac{\Delta E}{\mbox{1~GeV}} \,,
\label{delzeta}
\end{equation}
where $\Delta E$ is the difference in the energy release in different
decay channels. In the two-body $B$ decays of interest to us, $\Delta
E$ is always smaller than 1~GeV, so that $\Delta\zeta$ becomes a
parameter of order $1/m_b$. The corresponding variations of $\zeta$ are
of order a few per cent, which is small compared with the value of
$\zeta$ itself ($\zeta\approx 1/3$ for $\bar B\to D\pi$ decays).

To summarize this discussion, we repeat that in energetic two-body
decays of $B$ mesons the large-$N_c$ counting rules of QCD predict that
factorization with $a_1^{\rm eff}\approx 1$ works well for class~I
transitions, whereas class~II transitions can be described by a
phenomenological coefficient $a_2^{\rm eff}\approx c_2(m_b)+\zeta\,
c_1(m_b)$ with $\zeta=O(1/N_c)$. Based on an intuitive argument, we
have estimated that the process dependence of $\zeta$ is very mild,
since $\Delta\zeta\sim\Delta E/m_b$. Thus, we have provided a
theoretical basis for the phenomenological treatment of using a
factorization prescription in which the phenomenological parameter
$\zeta$ is taken to be the same for all energetic two-body decays of
$B$ mesons.

Let us briefly also discuss the case of charm decays. Here the
large-$N_c$ counting rules are different, because empirically
$L/N_c=O(1)$. Then, neglecting terms of order $1/N_c^2$, we find
instead of (\ref{a1a2appr}) the relations
\begin{equation}
   a_1^{\rm eff} \approx c_1(m_c) + \zeta'\,c_2(m_c) \,,\qquad
   a_2^{\rm eff} \approx c_2(m_c) + \zeta\,c_1(m_c) \,,
\end{equation}
with
\begin{equation}
   \zeta' \equiv {1\over N_c} + \varepsilon_8^{(DK,\pi)}(m_c)
   \,, \qquad
   \zeta \equiv {1\over N_c} + \varepsilon_8^{(D\pi,K)}(m_c) \,.
\end{equation}
In general, class~I and II transitions can no longer be described by
the same $\zeta$ parameter. However, we may again argue that the
expected process dependence of the $\zeta$ parameters is a mild one. In
(\ref{delzeta}) now appears the smaller charm-quark mass in the
denominator, but on the other hand the energy difference $\Delta E$ in
the numerator is smaller than in $B$ decays. As a consequence, we still
have $\Delta\zeta$ of order a few per cent; in particular, then, we
expect $\zeta'\approx\zeta$. That $\zeta$ in (\ref{a1a2appr}) should be
treated as a phenomenological parameter, rather than fixed to the value
$\zeta=1/3$ predicted by naive factorization, was the basis of the
approach of Bauer et al.~\cite{bsw87} (see also Ref.~53).
It
turns out that setting $\zeta\approx 0$ provides a successful
description of many two-body decays of $D$ mesons. To some extend, this
phenomenological ``rule of discarding $1/N_c$ terms'' can be understood
in the context of QCD sum rules~\cite{BlokS}, and using more formal
considerations based on the heavy-quark expansion~\cite{BlSh}. From the
above argument, it follows that if $\zeta\approx 0$ for one $D$ decay
mode, it is expected to be a small parameter also for other channels.

As a final comment, we note that whereas the concept of introducing
effective, process-dependent parameters $a_i^{\rm eff}$ works in most
two-body decays of $B$ and $D$ mesons, it has to be modified in decays
where the final state consists of two vector particles ($P\to V V$
decays). The polarization of the final-state particles in such
processes is very sensitive to non-factorizable contributions and
final-state interactions. The ratios between S-, P- and D-wave
amplitudes predicted in the factorization approximation are affected
since non-factorizable contributions to the amplitude will, in general,
have a different structure for different partial waves~\cite{Chen}. In
other words, the matrix elements describing the non-factorizable
contributions to the decay amplitudes are spin-dependent, thus
affecting the polarization of the final-state particles. Likewise,
final-state interactions are different for different partial waves.
They may change S- into D- waves even without changing the total decay
rate. A case of particular interest is the polarization of the $J/\psi$
particle in the decay $B\to\bar K^*J/\psi$. Most models predict a
longitudinal polarization of around 40\% (a model that we shall
introduce in Section~\ref{sec:formfa} predicts 48\%), whereas the
experimental world average is~\cite{honreview} $(78\pm 7)\%$. A recent
CLEO measurement, however, gives~\cite{Lewis} $(52\pm 7\pm 4)\%$.
Although a clear picture has not yet emerged, it is possible that the
first deviations from factorization predictions in $B$ decays will be
seen in polarization data. Indeed, it has been argued that
spin-dependent non-factorizable effects are necessary to understand the
data~\cite{Kamal} (see, however, also the discussion in Ref.~8).

\section{Final-State Interactions}
\label{sec:FSI}

In the conventional factorization approximation, the scattering of
the final-state particles off each other is neglected, and all
amplitudes are real (apart from an arbitrary common phase). Watson's
theorem, however, requires the amplitudes to have the same phases as
the corresponding scattering amplitudes~\cite{Watson}. Moreover, in
two-body decays, the final state must be of low angular momentum, and
we know that S- and P-wave scattering amplitudes in the GeV region
are, in general, inelastic.

Let us first consider a scattering eigenstate, i.e.\ an eigenstate of
the $S$ matrix. The phase of the amplitude for the decay into this
state, which is in general a linear combination of physical final
states, is identical to the corresponding scattering phase. But
final-state interactions will, in general, also affect the magnitude
of the scattering amplitudes. By definition, these interactions occur
in a space-time region where the final-state particles have already
been formed in their ground states, but are still strongly
interacting while recoiling from each other. Accordingly, for a very
energetic two-body decay with a pion in the final state, the ``colour
transparency argument'' of the previous section already excludes the
possibility of significant final-state interactions, since the light
quark-antiquark pair will have left the region of strong interaction
long before it hadronizes into a pion. In a weak decay involving small
recoil energies, however, the product $k R$ (particle momentum times
the radius of the strong-interaction region) is of the same order as
the scattering phase $\delta$, which is $O$(1). Therefore, in such a
decay rescattering will indeed change also the magnitude of the decay
amplitude. In $K\to 2\pi$ decays, for example, rescattering effects
lead to an enhancement of the $\Delta I=\frac{1}{2}$ amplitude by
$\approx 30$\% and to a reduction of the $\Delta I=\frac{3}{2}$
amplitude by $\approx 10$\%~\cite{diquarks}. Fortunately, in very
energetic processes such as exclusive $B$ decays, the phase $\delta$ is
negligible compared to $k R$, and we may safely assume that the
magnitude of the amplitude for decays into a scattering eigenstate
remains unchanged by final-state interactions. As a consequence, the
relation between the amplitudes $A_i$ for decays into final states $i$
(which are, in general, not scattering eigenstates) to the ``bare''
amplitudes $A_i^0$ is
\begin{equation}\label{S}
   A_i=\big( S^{1/2} \big)_{ij}\,A_j^0 \,,
\end{equation}
where $S$ denotes the strong-interaction $S$ matrix.

The $S$ matrix can redistribute the amplitudes into different
channels carrying the same quantum numbers, including those channels
which were not originally coupled to the weak process. To illustrate
the consequences of (\ref{S}) we consider, as an extreme example, a
two-channel $S$ matrix with maximal absorption ($S_{11}=S_{22}=0$):
\begin{equation}
   S=\left( \begin{array}{cc} 0 & e^{i\varphi}\\
           e^{i\varphi} & 0 \end{array} \right) \,.
\end{equation}
The square root of this matrix is
\begin{equation}
   S^{1/2} = {1\over\sqrt{2}}\,\exp\left\{
   i\,\left({\varphi\over 2} - {\pi\over 4} \right) \right\}\,
   \left(\begin{array}{cc} 1~ & i\\
         i~ & 1 \end{array} \right) \,.
\end{equation}
For a vanishing ``bare'' amplitude for decays into the second channel
($A_2^0=0$), the amplitude $A_1$ for decays into the first channel is
reduced by a factor $1/\sqrt{2}$, leading to a 50\% reduction in the
corresponding branching ratio. The second channel, in spite of not
being directly accessible through the weak decay, obtains an
amplitude of the same magnitude. Although this is an extreme example,
one should be aware of the fact that different decay channels
influence each other. In particular, decays with small branching
ratios may have been modified or even caused by a ``spill-over'' from
stronger modes.

Obviously, using the factorization approximation we can only attempt
to calculate the ``bare'' decay amplitudes $A_i^0$. However, summing
over all decay channels with the same conserved quantum numbers, the
uncertainties connected with final-state interactions drop out. The
reason is that, because of (\ref{S}) and the unitarity of the $S$
matrix, we have
\begin{equation}
  \sum_i |A_i|^2 = \sum_i |A_i^0|^2 \,,
\end{equation}
i.e.\ this sum of decay rates remains unaffected by final-state
interactions. By measuring the branching ratios of several related
decay channels, one can extract the magnitudes of the corresponding
isospin amplitudes as well as their relative phases~\cite{Ka86}. The
same isospin amplitudes can also be obtained from the ``bare'' decay
amplitudes calculated in the factorization approximation. Neglecting
inelastic rescattering, one can then extract the QCD coefficients
$a_1^{\rm eff}$ and $a_2^{\rm eff}$ from a comparison of measured and
calculated isospin amplitudes.

It is well-known that final-state interactions are important in $D$
decays; they must be included in the comparison of theoretical
predictions with data~\cite{firstedi}. For $B$ decays, on the other
hand, we expect smaller final-state interaction effects. Because of the
large energy released to the particles in the final state, the
magnitude of the isospin amplitudes should not be modified
significantly by final-state interactions. Moreover, there do not exist
any charm resonances in the $B$ meson mass region that could lead to
strong final-state interactions. As an example, consider the decays
$\bar B\to D\,\pi$. The four-quark operator $(\bar d u)(\bar c b)$,
which is part of the effective weak Hamiltonian relevant for $B$
decays, carries isospin quantum numbers $I=1$ and $I_3=+1$ and thus can
transform a $\bar B^0$ meson into a state with $I=\frac{1}{2}$ or
$I=\frac{3}{2}$, while the $B^-$ can only decay into final states with
$I=\frac{3}{2}$. Accordingly, we define the following (complex) isospin
amplitudes:
\begin{eqnarray}\label{isoamp}
   A_{1/2} &=& \langle D\pi;I=\textstyle\frac{1}{2}|\,
    H_{\rm eff}\,|\bar B^0\rangle \nonumber\\
   A_{3/2} &=& \langle D\pi;I=\textstyle\frac{3}{2}|\,
    H_{\rm eff}\,|\bar B^0\rangle
   = \displaystyle{\frac{1}{\sqrt{3}}}\,
    \langle D\pi;I=\textstyle\frac{3}{2}|\,H_{\rm eff}\,|B^-\rangle \,.
\end{eqnarray}
They are related with the physical decay amplitudes through
\begin{eqnarray}\label{isozerl}
   A(\bar B^0\to D^+\pi^-) &=& \sqrt{1\over 3}\,A_{3/2}
    + \sqrt{2\over 3}\,A_{1/2} \nonumber\\
   A(\bar B^0\to D^0\pi^0) &=& \sqrt{2\over 3}\,A_{3/2}
    - \sqrt{1\over 3}\,A_{1/2} \nonumber\\
   A(B^-\to D^0\pi^-) &=& \sqrt{3}\,A_{3/2} \,.
\end{eqnarray}
Although the phases of the measured decay amplitudes are unknown, we
can extract from (\ref{isozerl}) the isospin amplitudes as well as
their relative phase:
\begin{eqnarray}\label{isosol}
   |A_{1/2}|^2 &=& |A(\bar B^0\to D^+\pi^-)|^2
    + |A(\bar B^0\to D^0\pi^0)|^2 \nonumber\\
   &&\mbox{}- {1\over 3}\,|A(B^-\to D^0\pi^-)|^2 \nonumber\\
   |A_{3/2}|^2 &=& {1\over 3}\,|A(B^-\to D^0\pi^-)|^2 \nonumber\\
   \cos(\delta_{3/2}-\delta_{1/2}) &=&
    {3|A(\bar B^0\to D^+\pi^-)|^2 - 2|A_{1/2}|^2 - |A_{3/2}|^2
     \over 2\sqrt{2} |A_{1/2}||A_{3/2}|} \,.
\end{eqnarray}
Let us assume that the magnitudes of the three physical amplitudes
had been measured. We could then use these relations to determine the
magnitudes of the two isospin amplitudes as well as their relative
phase. In order to compare the results with the amplitudes calculated
in the factorization approximation, we must assume that the magnitude
of the isospin amplitudes remains unaffected by final-state
interactions. In other words, the contribution of inelastic scattering
into or from other channels has to be negligible. Then neglecting
final-state interactions simply amounts to neglecting the relative
phase shift between the different isospin amplitudes, and we can
directly compare the theoretical predictions obtained for $|A_{1/2}|$
and $|A_{3/2}|$ with the experimental values for these quantities.

\begin{table}
\caption{\label{tab:deltaiso}
Upper limits for the relative phase shifts between the two isospin
amplitudes in some exclusive hadronic $B$ decays}
\vspace{0.4cm}
\begin{center}
\begin{tabular}{|l|cc|}\hline
\rule[-0.2cm]{0cm}{0.6cm}
Ratio & CLEO (90\% CL) & $|\delta_{3/2}-\delta_{1/2}|$ \\[0.1cm]
\hline
B$(\bar B^0\to D^0\pi^0)$/B$(B^-\to D^0\pi^-)$
\rule[-0.2cm]{0cm}{0.6cm} & $<0.07$ & $<36^\circ$ \\[0.1cm]
B$(\bar B^0\to D^0\varrho^0)$/B$(B^-\to D^0\varrho^-)$ & $<0.05$ &
 $<30^\circ$ \\[0.1cm]
B$(\bar B^0\to D^{*0}\pi^0)$/B$(B^-\to D^{*0}\pi^-)$ & $<0.13$ &
 $<53^\circ$ \\[0.1cm]
B$(\bar B^0\to D^{*0}\varrho^0)$/B$(B^-\to D^{*0}\varrho^-)$ &
 $<0.07$ & $<36^\circ$ \\[0.1cm]
\hline
\end{tabular}
\end{center}
\end{table}

Unfortunately, in $B$ decays complete measurements of the class of
decay amplitudes relevant to the isospin analysis described above are
not yet available. Still, we can use existing data to obtain upper
bounds on the relative phase shifts in some interesting cases. In our
example, it is the amplitude $A(\bar B^0\to D^0\pi^0)$ that has not yet
been measured. However, from the current upper limit for the
corresponding branching ratio we can derive an upper limit for the
relative phase shift $|\delta_{3/2}-\delta_{1/2}|$. To this end, we
represent the second relation in (\ref{isozerl}) as a triangle in the
complex plane. An elementary geometrical argument shows that the angle
between $A_{1/2}$ and $A_{3/2}$ is maximal when there is a right angle
between $A_{1/2}$ and $A(\bar B^0\to D^0\pi^0)$. Hence
\begin{equation}\label{deltalimit}
   \sin^2(\delta_{3/2}-\delta_{1/2})
   \le {9\over 2}\,{\tau(B^-)\over\tau(\bar B^0)}\,
   {\mbox{B}(\bar B^0\to D^0\pi^0)\over\mbox{B}(B^-\to D^0\pi^-)} \,.
\end{equation}
Similar inequalities can be derived for the phase shifts in other decay
channels. Using then experimental upper limits for colour-suppressed
$B$ decay modes obtained by the CLEO
Collaboration~\cite{honreview,Rodri}, as well as $\tau(B^-)/\tau(\bar
B^0)=1.06\pm 0.04$ for the lifetime ratio~\cite{Rich}, we obtain the
results shown in Table~\ref{tab:deltaiso}. The next generation of CLEO
data is likely to produce still smaller upper bounds for the various
phase shifts, but even the current limits are a strong indication that
final-state interactions in hadronic two-body decays of $B$ mesons are
much less important than in the corresponding decays of $D$ mesons. We
will therefore neglect them in our analysis.

\section{Decay Constants}
\label{sec:decay}

The evaluation of factorized amplitudes requires the knowledge of
meson decay constants and hadronic form factors of current matrix
elements. We shall discuss first what is known about decay constants.
For a pseudoscalar meson $P=(q_1\bar q_2)$, we define
\begin{equation}
   \langle\,0\,|\,\bar q_2\gamma_\mu\gamma_5 q_1\,|P(p)\rangle
   = i f_P p_\mu \,.
\end{equation}
The decay constant $f_V$ of a vector meson $P=(q_1\bar q_2)$ is
defined by~\footnote{For the neutral mesons $\varrho^0$ and $\omega$,
we use the current $\bar u\gamma_\mu u$ on the left-hand side of
(\protect\ref{fdef}) and divide the right-hand side by $\sqrt{2}$.
With this definition and under the assumption of isospin symmetry,
the decay constants of neutral and charged $\varrho$ mesons have the
same numerical value.}
\begin{equation}\label{fdef}
   \langle\,0\,|\,\bar q_2\gamma_\mu q_1\,|V(\epsilon)\rangle
   = \epsilon_\mu m_V f_V \,.
\end{equation}

The experimental values of the decay constants of the charged pion
and kaon, as obtained from their leptonic decays
$P^+\to\ell^+\nu_\ell\,(\gamma)$, are~\cite{PDG}
\begin{equation}
   f_\pi = (130.7\pm 0.37)~\mbox{MeV} \,, \qquad
   f_K = (159.8\pm 1.47)~\mbox{MeV} \,.
\label{fpifK}
\end{equation}
For charm mesons, the uncertainties in the experimental values of the
leptonic decay rates are much larger. For the $D^+$ meson, only an
upper bound can be deduced from the published data~\cite{fDlimit}:
\begin{equation}
   f_D < 310~\mbox{MeV}~~\mbox{(90\% CL)} \,.
\end{equation}
Four groups (WA75~\cite{aoki}, CLEO~\cite{franz}, BES~\cite{fDs3}, and
E653~\cite{E653}) have measured the $D_s^+\to\mu^+\nu_\mu$ branching
ratio, from which the decay constant $f_{D_s}$ can be extracted. The
average value is~\cite{Rich}
\begin{equation}
   f_{D_s} = (241\pm 37)~\mbox{MeV} \,.
\label{fDs}
\end{equation}
Later, we shall compare this result with an independent determination
of $f_{D_s}$ from non-leptonic decays.

The decay constants of light charged mesons can also be obtained from
the 1-prong hadronic decays of the $\tau$ lepton. Denoting the relevant
CKM-matrix element by $V_{ij}$, and neglecting radiative corrections,
we write the corresponding decay width as
\begin{equation}\label{tauwidth}
   \Gamma(\tau^-\to M^-\nu_\tau) = {m_\tau^3\over 16\pi}\,G_F^2\,
   |V_{ij}|^2\,f_M^2\,\left( 1 - {m_M^2\over m_\tau^2} \right)^2
   \left( 1 + b_M {m_M^2\over m_\tau^2} \right) \,,
\end{equation}
with $b_P=0$ for pseudoscalar and $b_V=2$ for (axial) vector mesons. We
use $m_\tau=1777$~MeV for the mass of the $\tau$ lepton and
$\tau_\tau=291.0\pm 1.5$~fs for its lifetime~\cite{PDG}. Moreover, we
use the following branching ratios for $\tau^-\to M^-
\nu_\tau$~\cite{PDG,Evans}: $(11.31\pm 0.15)\%~(\pi)$, $(0.69\pm
0.03)\%~(K)$, $(24.94\pm 0.16)\%~(\varrho)$, $(1.25\pm 0.08)\%~(K^*)$,
and $(17.65\pm 0.32)\%~(a_1)$. This leads to the values of the decay
constants shown in Table~\ref{tab:fM}. Only experimental errors are
quoted. The results for $f_\pi$ and $f_K$ are in excellent agreement
with those given in (\ref{fpifK}).

\begin{table}
\caption{\label{tab:fM}
Decay constants of light charged mesons extracted from hadronic $\tau$
decays}
\vspace{0.4cm}
\begin{center}
\begin{tabular}{|c|ccccc|}\hline
\rule[-0.2cm]{0cm}{0.6cm} $M$ \rule[-0.2cm]{0cm}{0.6cm} &
 $\pi$ & $K$ & $\varrho$ & $K^*$ & $a_1$ \\
\hline
\rule[-0.2cm]{0cm}{0.6cm} $f_M$~[MeV] \rule[-0.2cm]{0cm}{0.6cm} &
 $134\pm 1$ & $158\pm 3$ & $208\pm 1$ & $214\pm 7$ & $229\pm 10$ \\
\hline
\end{tabular}
\end{center}
\end{table}

The decay constants of neutral vector mesons can be extracted from
their electromagnetic decay width, using
\begin{equation}
   \Gamma(V^0\to e^+ e^-) = {4\pi\over 3}\,{\alpha^2\over m_V}\,
   f_V^2\,c_V \,,
\end{equation}
where $c_V$ are factors related to the electric charge of the quarks
that make up the vector meson. From the measured widths~\cite{PDG}, we
obtain the results shown in Table~\ref{tab:fV}. The errors reflect the
uncertainty in the experimental data only. The value for $f_\varrho$
obtained in this way is somewhat larger than that derived from $\tau$
decays. In our analysis below, we take $f_\varrho=210$~MeV. In all
other cases, we take the central values shown in the Tables.

\begin{table}
\caption{\label{tab:fV}
Decay constants of neutral vector mesons extracted from their
electromagnetic decays}
\vspace{0.4cm}
\begin{center}
\begin{tabular}{|c|ccccc|}\hline
\rule[-0.2cm]{0cm}{0.6cm} $V$ \rule[-0.2cm]{0cm}{0.6cm} &
 $\varrho$ & $\omega$ & $\phi$ & $J/\psi$ & $\psi(\mbox{2S})$ \\
\hline
\rule[-0.2cm]{0cm}{0.6cm} $c_V$ \rule[-0.2cm]{0cm}{0.6cm} &
 1/2 & 1/18 & 1/9 & 4/9 & 4/9 \\[0.1cm]
$f_V$~[MeV] & $216\pm 5$ & $195\pm 3$ & $237\pm 4$ & $405\pm 14$ &
 $282\pm 14$ \\[0.1cm]
\hline
\end{tabular}
\end{center}
\end{table}

We note for completeness that the decay constants of light neutral
pseudoscalar mesons can be extracted from the two-photon decay
$P^0\to\gamma\gamma$. Averaging the results reported by the
TPC/$2\gamma$~\cite{feta1} and CELLO~\cite{feta2} Collaborations, we
find
\begin{equation}
   f_\eta = (131\pm 6)~\mbox{MeV} \,, \qquad
   f_{\eta'} = (118\pm 5)~\mbox{MeV} \,.
\end{equation}

Decay constants not yet known experimentally are left as free
parameters in our expressions for the branching ratios. However, we
will often need an estimate for the decay constants of charm mesons,
such as $D$ and $D^*$. In the heavy-quark limit ($m_c\to\infty$), spin
symmetry predicts that $f_D=f_{D^*}$, and most theoretical predictions
indicate that symmetry-breaking corrections enhance the ratio
$f_{D^*}/f_D$ by 10--20\% \cite{review}. Hence, we take as our central
values
\begin{eqnarray}\label{fD}
   f_D &=& 200~{\rm MeV} \,,\qquad f_{D^*} = 230~{\rm MeV} \,,
    \nonumber\\
   f_{D_s} &=& 240~{\rm MeV} \,,\qquad f_{D_s^*} = 275~{\rm MeV} \,.
\end{eqnarray}
The value for $f_D$ lies in the ball park of most theoretical
predictions, whereas that for $f_{D_s}$ corresponds to the central
experimental result quoted in (\ref{fDs}).

\section{Transition Form Factors}
\label{sec:formfa}

The most important ingredient of factorized decay amplitudes are the
hadronic form factors defined in terms of the covariant decomposition
of hadronic matrix elements of current operators. In particular, we
need matrix elements of the type
\begin{equation}
   \langle M|\,\bar q\gamma_\mu(1-\gamma_5) b\,|B\rangle \,,
\end{equation}
where $M$ is a pseudoscalar or vector meson with mass $m_M$. There are
two form factors ($F_0$, $F_1$) describing the transition to a
pseudoscalar particle, and four form factors ($V$, $A_0$, $A_1$, $A_2$)
for the transition to a vector particle. The definitions of these form
factors are given in the Appendix. Conventionally, they are written as
functions of the invariant momentum transfer $q^2$. To obtain reliable
theoretical predictions for the form factors is the main obstacle in
the analysis of hadronic weak decays, once the factorization hypothesis
is accepted.

In the case of the heavy-to-heavy transitions $\bar B\to D$ and $\bar
B\to D^*$, heavy-quark symmetry implies simple relations between the
various transition form factors. In the heavy-quark limit, they
read~\cite{Rieck}
\begin{eqnarray}\label{HQLff}
   F_1(q^2) &=& V(q^2) = A_0(q^2) = A_2(q^2)
    = {m_B + m_M\over 2\sqrt{m_B m_M}}\,\xi(w) \,, \nonumber\\
   F_0(q^2) &=& A_1(q^2) = {2\sqrt{m_B m_M}\over m_B + m_M}\,
    {w+1\over 2}\,\xi(w) \,,
\end{eqnarray}
where
\begin{equation}
   w = v_B\cdot v_M = \frac{m_B^2 + m_M^2 - q^2}{2 m_B m_M}
\end{equation}
is the product of the velocities of the two mesons, and the universal
(mass-independent) function $\xi(w)$ is called the Isgur-Wise form
factor~\cite{Isgu,Falk}. This function is normalized to unity at the
kinematic point $w=1$, where the two mesons have a common rest frame.
For the realistic case of finite heavy-quark masses, the above
relations are modified by corrections that break heavy-quark symmetry.
They can be analysed in a systematic way using the heavy-quark
effective theory \cite{review}$^-$\cite{vcb}. This is discussed in
detail in the article by one of us (M.~Neubert) in this volume. As a
consequence, it has become possible to extract the $\bar B\to D^{(*)}$
form factors relevant for all class~I transitions considered in this
article from semileptonic decay data with good precision and in an
essentially model-independent way. The results are compiled in
Ref.~57.
The use of heavy-quark symmetry constitutes a
significant improvement compared with earlier estimates of non-leptonic
amplitudes, which were based on model calculations of the relevant form
factors.

Unfortunately, there has not been similar progress in the calculation
of hadronic current matrix elements between heavy and light mesons. For
these transitions, we still must rely on the results obtained using
some phenomenological model. One such model (below referred to as the
NRSX model) was introduced in Ref.~57,
where we used the
overlap integrals of the BSW model~\cite{ba85,bsw87} to obtain the form
factors at zero momentum transfer, and then proposed a specific ansatz
for their $q^2$ dependence motivated by pole dominance and the
relations in (\ref{HQLff}). Another model~\cite{Casa}, which we shall
not discuss in detail, has been used by Deandrea et al.~\cite{Dean} to
perform a global analysis of non-leptonic decays comparable to ours.

In order to get an idea about the unavoidable amount of model
dependence, which will affect our predictions for class~II decays of
$B$ mesons, we will consider an alternative to the NRSX model. The main
purpose here is not to provide a new approach that is more
sophisticated; on the contrary, we will make several strong assumptions
to obtain a model as simple as possible. Yet, we have checked that it
agrees, within reasonable limits, with the available data on
semileptonic decays, as well as with most theoretical predictions. We
thus feel confident that this model may also be used advantageously to
obtain rough estimates for other processes, such as rare $B$ decays and
decays of $B_s$ mesons.

In the case of heavy-to-light transitions, the symmetry relations in
(\ref{HQLff}) are no longer valid. We account for this by introducing,
for each form factor, a function $\xi_i(w)$ replacing the Isgur-Wise
function. There is, however, still a connection between various form
factors provided by kinematic constraints at zero momentum transfer
($q^2=0$), corresponding to
\begin{equation}
   w=w_{\rm max}=\frac{m_B^2+m_M^2}{2 m_B m_M} \,.
\end{equation}
There, the form factors satisfy the relations
\begin{eqnarray}
   F_1(0) &=& F_0(0) \,, \nonumber\\
   A_0(0) &=& \frac{m_B+m_M}{2 m_M}\,A_1(0)
    - \frac{m_B-m_M}{2 m_M}\,A_2(0) \,.
\label{kinrel}
\end{eqnarray}
The normalization condition for the Isgur-Wise function is replaced by
the inequalities
\begin{equation}
   \xi_{F_0}(1) \le 1 \,, \qquad \xi_{A_1}(1) \le 1 \,,
\label{relnorm}
\end{equation}
which follow from equal-time commutation relations for a heavy $b$
quark. For an estimate of the functions $\xi_i(w)$ we shall adopt a
simple pole model, which apart from the masses of some resonances has
no tunable parameters. This model reproduces the relations
(\ref{HQLff}) in the heavy-quark limit. It will serve for an immediate
estimate of all form factors in the physical region of $q^2$. As a
starting point, we write $\xi_i(w)=\sqrt{2/(w+1)}\,h_i(w)$, where
$h_i(w)$ will be simple monopole form factors. In the heavy-quark
limit, the prefactor $\sqrt{2/(w+1)}$ can be shown to yield an upper
bound for the Isgur-Wise function~\cite{Rieck}. For heavy-to-light form
factors, on the other hand, this factor ensures that $\xi_i(w)\sim
w^{-3/2}$ for large $w$, in accordance with the scaling rules obtained
in Ref.~97.
In the next step we make the strong assumption
that the symmetry relations in (\ref{HQLff}) still hold for
heavy-to-light form factors, but only at the point $q^2=0$. In other
words, we assume that all the $h_i(w_{\rm max})$ are equal. This is a
minimal ansatz that preserves the kinematic relations in
(\ref{kinrel}). For the functions $h_i(w)$ with $i=F_1$, $V$, $A_0$ and
$A_2$, we take
\begin{equation}
   h_i(w) = N\,\frac{w_{\rm max}-w_i}{w-w_i} \,,
\label{hidef}
\end{equation}
where $N$ is a common normalization factor. Note that, according to its
definition, the value of $w_{\rm max}$ in $h_{F_1}$ is different from
that in the other three cases. The locations of the poles are at
\begin{equation}
   w_i = \frac{m_B^2+m_M^2-M_i^2}{2 m_B m_M} \,,
\end{equation}
where $M_i$ denotes the mass of the nearest resonance with the
appropriate spin-parity quantum numbers. Most of these masses are known
experimentally; some others (such as the masses of the $B_c$ and
$B_c^*$ mesons) are taken from potential models. To obtain the masses
of the $1^+$ resonances, we simply add 400~MeV to the masses of the
corresponding $1^-$ states, as suggested by the spectroscopy of the
light and charm mesons, and use the same values for the $A_1$ and $A_2$
form factors.

It remains to find an ansatz for $h_{F_0}(w)$ and $h_{A_1}(w)$. For
simplicity, we choose to normalize these functions to unity at $w=1$,
corresponding to a complete overlap of the initial and final states at
the same velocity [cf.~(\ref{relnorm})]. Moreover, we have to satisfy
the condition $h_{F_0}(w_{\rm max}) = h_{A_1}(w_{\rm max}) = N$. Thus,
we take (for $j=F_0$, $A_1$)
\begin{equation}
   h_j(w) = \displaystyle\frac{1}{1+ r\,\displaystyle
    \frac{w-1}{w_{\rm max}-1}} \,,
\end{equation}
where $N=1/(1+r)$ is still a free parameter. To fix it, we try to mimic
the effect of the physical axial vector ($1^+$) pole on the form factor
$A_1(w)$ near $w=1$. In order to have full correspondence with the pole
structure of the other form factors in (\ref{hidef}), the factor
$\frac 12(w+1)$ in (\ref{HQLff}) has to be included. Therefore, we
require that the derivative of the product $\frac 12(w+1)\,h_{A_1}(w)$
at $w=1$ be equal to the derivative of a single-pole form factor with
the physical pole mass $M_{A_1}$. The result is
\begin{equation}
   r = \frac{(m_B-m_V)^2}{4 m_B m_V} \left( 1
   + \frac{4 m_B m_V}{M_{A_1}^2 - (m_B - m_V)^2} \right) \,,
\end{equation}
where $V$ is a vector meson. The same value of $r$ is then taken for
$h_{F_0}(w)$. With these simple assumptions, all form factors are
determined.

A few tests of this model may be quoted here: The branching ratios for
the semileptonic decays $\bar B\to D^*\ell\,\bar\nu$ and $\bar B\to
D\,\ell\,\bar\nu$ are found to be 5.3\% and 1.8\%, respectively. The
corresponding experimental values are~\cite{PDG} $(4.64\pm 0.26)\%$ and
$(1.8\pm 0.4)\%$. In the semileptonic decay $\bar B\to
\varrho\,\ell\,\bar\nu$, the values for the form factors at $q^2=0$ are
$V=A_1=A_2=0.26$, in accordance with the predictions $V=0.35\pm 0.07$,
$A_1=0.27\pm 0.05$ and $A_2=0.28\pm 0.05$ obtained using QCD sum
rules~\cite{BaBraun}. The expected branching ratio for this decay is
$25.5|V_{ub}|^2$, which when combined with the experimental
result~\cite{Bpirho} $B(\bar B\to\varrho\,\ell\,\bar\nu)=
(2.5_{-0.9}^{+0.8})\times 10^{-4}$ yields $|V_{ub}|=(3.1\pm 0.5)\times
10^{-3}$. The calculated $\varrho/\pi$ ratio in exclusive semileptonic
$B$ decays is 2.2, as compared with the experimental value
of~\cite{Bpirho} $1.4\pm 0.6$. Finally, comparing the calculated $\bar
B\to\pi$ form factor $F_1(q^2)$ at threshold with the $B^*$-pole
structure at this point~\cite{us,Kambor} gives
$g_{BB^*\pi}=0.54\times(0.2\,\mbox{GeV}/f_{B^*})$ for the $B B^*\pi$
coupling constant, which is of the expected order.

\boldmath
\section{Theoretical Predictions, Comparison with Experiment and
Determination of $a_1^{\rm eff}$ and $a_2^{\rm eff}$}
\unboldmath
\label{sec:comparison}

We first discuss the results obtained using the NRSX
model~\cite{firstedi}. To obtain predictions for the branching ratios
from the factorized decay amplitudes, we use $\tau(\bar B^0)=1.55\pm
0.04$~ps and $\tau(B^-)=1.65\pm 0.04$~ps for the $B$-meson
lifetimes~\cite{Rich}, and $V_{cb}=0.039\pm 0.002$ from (\ref{Vcbval}).
The small errors in these quantities, as well as uncertainties in the
decay constants of light mesons, are neglected. Our predictions for the
branching ratios of the dominant non-leptonic two-body decays of $B$
mesons are given in Tables~\ref{tab:B0decays} and \ref{tab:Bmidecays}.
The QCD coefficients $a_1^{\rm eff}$ and $a_2^{\rm eff}$ (for
simplicity called $a_1$ and $a_2$ in the tables), as well as the
unknown decay constants of charm mesons, have been left as parameters
in the expressions for the branching ratios. For comparison, we show
the world average experimental results for the branching ratios, as
recently compiled in the review article in Ref.~102.
They are
dominated by the CLEO~II measurements. The quoted upper limits for the
colour-suppressed $\bar B^0$ decay modes also include recent CLEO data
reported in Ref.~86.
In Tables~\ref{tab:B0new}--\ref{tab:Bminew}, we show for comparison the
predictions obtained using the simple form-factor model described in
the previous section, including some new decay channels which had not
been considered in Ref.~57.
In the theoretical expression
for the class~III decay $B^-\to D^{*0} a_1^-$, the term proportional to
$a_2^{\rm eff}$ requires the knowledge of the current matrix element
between the $B$ meson and the pseudovector meson $a_1$. We use the
axial current as an interpolating field for the $a_1$ particle and
employ chiral symmetry to relate the corresponding form factors to
those of the $B^-\to\varrho^-$ matrix element.

\begin{table}
\caption{\label{tab:B0decays}
Branching ratios of non-leptonic $\bar B^0$ decays (in \%) obtained in
the NRSX model~\protect\cite{firstedi}. In the third column, the
factors containing not yet known decay constants have been suppressed.
The last column shows the world average experimental
results~\protect\cite{newrev,Rodri}. Upper limits are given at the 90\%
confidence level}
\vspace{0.4cm}
\begin{center}
\begin{tabular}{|l|cc|c|}
\hline
$\bar B^0$ Modes \rule{0cm}{0.4cm} & NRSX Model &
 $a_1^{\rm eff}=1.08$ & Experimental \\
 & & $a_2^{\rm eff}=0.21$ & Average \\[0.1cm]
\hline
Class I \rule[-0.2cm]{0cm}{0.6cm} & & & \\
\hline
$D^+\pi^-$ \rule{0cm}{0.4cm} & 0.257 $a_1^2$ & 0.30 &
 $0.31\pm 0.04\pm 0.02$ \\
$D^+ K^-$ & 0.020 $a_1^2$ & 0.02 & \\
$D^+ D^-$ & 0.031 $a_1^2\,(f_D/200)^2$ & 0.04 & \\
$D^+ D_s^-$ & 0.879 $a_1^2\,(f_{D_s}/240)^2$ & 1.03 &
 $0.74\pm 0.22\pm 0.18$ \\
$D^+\varrho^-$ & 0.643 $a_1^2$ & 0.75 & $0.84\pm 0.16\pm 0.07$ \\
$D^+ K^{*-}$ & 0.035 $a_1^2$ & 0.04 & \\
$D^+ D^{*-}$ & 0.030 $a_1^2\,(f_{D^*}/230)^2$ & 0.03 & \\
$D^+ D^{*-}_s$ & 0.817 $a_1^2\,(f_{D_s^*}/275)^2$ & 0.95 &
 $1.14\pm 0.42\pm 0.28$ \\
$D^+ a_1^-$ & 0.719 $a_1^2$ & 0.84 & \\
$D^{*+}\pi^-$ & 0.247 $a_1^2$ & 0.29 & $0.28\pm 0.04\pm 0.01$ \\
$D^{*+} K^-$ & 0.019 $a_1^2$ & 0.02 & \\
$D^{*+} D^-$ & 0.022 $a_1^2\,(f_D/200)^2$ & 0.03 & \\
$D^{*+} D_s^-$ & 0.597 $a_1^2\,(f_{D_s}/240)^2$ & 0.70 &
 $0.94\pm 0.24\pm 0.23$ \\
$D^{*+}\varrho^-$ & 0.727 $a_1^2$ & 0.85 & $0.73\pm 0.15\pm 0.03$ \\
$D^{*+} K^{*-}$ & 0.042 $a_1^2$ & 0.05 & \\
$D^{*+} D^{*-}$ & 0.072 $a_1^2\,(f_{D^*}/230)^2$ & 0.08 & \\
$D^{*+} D_s^{*-}$ & 2.097 $a_1^2\,(f_{D_s^*}/275)^2$ & 2.45 &
 $2.00\pm 0.54\pm 0.49$ \\
$D^{*+} a_1^-$ & 1.037 $a_1^2$ & 1.21 & $1.27\pm 0.30\pm 0.05$
\\[0.1cm]
\hline
Class II \rule[-0.2cm]{0cm}{0.6cm} & & & \\
\hline
$\bar K^0 J/\psi$ \rule{0cm}{0.4cm} & 2.262 $a_2^2$ & 0.10 &
 $0.075\pm 0.021$ \\
$\bar K^0 \psi(\mbox{2S})$ & 1.051 $a_2^2$ & 0.05 & $<0.08$ \\
$\bar K^{*0} J/\psi$ & 3.645 $a_2^2$ & 0.16 & $0.153\pm 0.028$ \\
$\bar K^{*0} \psi(\mbox{2S})$ & 1.939 $a_2^2$ & 0.09 &
 $0.151\pm 0.091$ \\
$\pi^0 D^0$ & 0.164 $a_2^2\,(f_D/200)^2$ & 0.007 & $<0.033$ \\
$\pi^0 D^{*0}$ & 0.230 $a_2^2\,(f_{D^*}/230)^2$ & 0.010 & $<0.055$ \\
$\varrho^0 D^0$ & 0.111 $a_2^2\,(f_D/200)^2$ & 0.005 & $<0.055$ \\
$\varrho^0 D^{*0}$ & 0.240 $a_2^2\,(f_{D^*}/230)^2$ & 0.011 & $<0.117$
 \\[0.1cm]
\hline
\end{tabular}
\end{center}
\end{table}

\begin{table}
\caption{\label{tab:Bmidecays}
Branching ratios (in \%) of non-leptonic $B^-$ decays in the NRSX
model}
\vspace{0.4cm}
\begin{center}
\begin{tabular}{|l|cc|c|}
\hline
$B^-$ Modes \rule{0cm}{0.4cm} & NRSX Model & $a_1^{\rm eff}=1.08$ &
 Experimental \\
 & & $a_2^{\rm eff}=0.21$ & Average \\[0.1cm]
\hline
Class I \rule[-0.2cm]{0cm}{0.6cm} & & & \\
\hline
$D^0 D^-$ \rule{0cm}{0.4cm} & 0.033 $a_1^2\,(f_D/200)^2$ &
 0.04 & \\
$D^0 D_s^-$ & 0.938 $a_1^2\,(f_{D_s}/240)^2$ & 1.09 &
 $1.36\pm 0.28\pm 0.33$ \\
$D^0 D^{*-}$ & 0.032 $a_1^2\,(f_{D^*}/230)^2$ & 0.04 & \\
$D^0 D^{*-}_s$ & 0.873 $a_1^2\,(f_{D_s^*}/275)^2$ & 1.02 &
 $0.94\pm 0.31\pm 0.23$ \\
$D^{*0} D^-$ & 0.023 $a_1^2\,(f_D/200)^2$ & 0.03 & \\
$D^{*0} D_s^-$ & 0.639 $a_1^2\,(f_{D_s}/240)^2$ & 0.75 &
 $1.18\pm 0.36\pm 0.29$ \\
$D^{*0} D^{*-}$ & 0.077 $a_1^2\,(f_{D^*}/230)^2$ & 0.09 & \\
$D^{*0} D^{*-}_s$ & 2.235 $a_1^2\,(f_{D_s^*}/275)^2$ & 2.61 &
 $2.70\pm 0.81\pm 0.66$ \\[0.1cm]
\hline
Class II \rule[-0.2cm]{0cm}{0.6cm} & & & \\
\hline
$K^- J/\psi$ \rule{0cm}{0.4cm} & 2.411 $a_2^2$ & 0.11 &
 $0.102\pm 0.014$ \\
$K^-\psi(\mbox{2S})$ & 1.122 $a_2^2$ & 0.05 & $0.070\pm 0.024$ \\
$K^{*-} J/\psi$ & 3.886 $a_2^2$ & 0.17 & $0.174\pm 0.047$ \\
$K^{*-} \psi(\mbox{2S})$ & 2.070 $a_2^2$ & 0.09 & $<0.30$ \\[0.1cm]
\hline
Class III \rule[-0.2cm]{0cm}{0.6cm} & & & \\
\hline
$D^0\pi^-$ \rule{0cm}{0.4cm} & $0.274\left[a_1+1.127\,a_2\,
 (f_D/200)\right]^2$ & 0.48 & $0.50\pm 0.05\pm 0.02$ \\
$D^0\varrho^-$ & $0.686\left[a_1+0.587\,a_2\,(f_D/200)\right]^2$ &
 0.99 & $1.37\pm 0.18\pm 0.05$ \\
$D^{*0}\pi^-$ & $0.264\left[a_1+1.361\,a_2\,(f_{D^*}/230)\right]^2$ &
 0.49 & $0.52\pm 0.08\pm 0.02$ \\
$D^{*0}\varrho^-$ & $0.775\,[a_1^2+0.661\,a_2^2\,(f_{D^*}/230)^2$ &
 1.19 & $1.51\pm 0.30\pm 0.02$ \\
 & $\phantom{0.775[} \mbox{}+ 1.518\,a_1 a_2\,(f_{D^*}/230)]$ &
 & \\[0.1cm]
\hline
\end{tabular}
\end{center}
\end{table}

\begin{table}
\caption{\label{tab:B0new}
Branching ratios (in \%) of class~I non-leptonic $\bar B^0$ decays in
the new model described in Section~\protect\ref{sec:formfa}}
\vspace{0.4cm}
\begin{center}
\begin{tabular}{|l|cc|c|}
\hline
$\bar B^0$ Modes \rule{0cm}{0.4cm} & New Model &
 $a_1^{\rm eff}=0.98$ & Experimental \\
 & & $a_2^{\rm eff}=0.29$ & Average \\[0.1cm]
\hline
Class I \rule[-0.2cm]{0cm}{0.6cm} & & & \\
\hline
$D^+\pi^-$ \rule{0cm}{0.4cm} & 0.318 $a_1^2$ & 0.30 &
 $0.31\pm 0.04\pm 0.02$ \\
$D^+ K^-$ & 0.025 $a_1^2$ & 0.02 & \\
$D^+ D^-$ & 0.037 $a_1^2\,(f_D/200)^2$ & 0.03 & \\
$D^+ D_s^-$ & 1.004 $a_1^2\,(f_{D_s}/240)^2$ & 0.96 &
 $0.74\pm 0.22\pm 0.18$ \\
$D^+\varrho^-$ & 0.778 $a_1^2$ & 0.75 & $0.84\pm 0.16\pm 0.07$ \\
$D^+ K^{*-}$ & 0.041 $a_1^2$ & 0.04 & \\
$D^+ D^{*-}$ & 0.032 $a_1^2\,(f_{D^*}/230)^2$ & 0.03 & \\
$D^+ D^{*-}_s$ & 0.830 $a_1^2\,(f_{D_s^*}/275)^2$ & 0.80 &
 $1.14\pm 0.42\pm 0.28$ \\
$D^+ a_1^-$ & 0.844 $a_1^2$ & 0.81 & \\
$D^{*+}\pi^-$ & 0.296 $a_1^2$ & 0.28 & $0.28\pm 0.04\pm 0.01$ \\
$D^{*+} K^-$ & 0.022 $a_1^2$ & 0.02 & \\
$D^{*+} D^-$ & 0.023 $a_1^2\,(f_D/200)^2$ & 0.02 & \\
$D^{*+} D_s^-$ & 0.603 $a_1^2\,(f_{D_s}/240)^2$ & 0.58 &
 $0.94\pm 0.24\pm 0.23$ \\
$D^{*+}\varrho^-$ & 0.870 $a_1^2$ & 0.84 & $0.73\pm 0.15\pm 0.03$ \\
$D^{*+} K^{*-}$ & 0.049 $a_1^2$ & 0.05 & \\
$D^{*+} D^{*-}$ & 0.085 $a_1^2\,(f_{D^*}/230)^2$ & 0.08 & \\
$D^{*+} D_s^{*-}$ & 2.414 $a_1^2\,(f_{D_s^*}/275)^2$ & 2.32 &
 $2.00\pm 0.54\pm 0.49$ \\
$D^{*+} a_1^-$ & 1.217 $a_1^2$ & 1.16 & $1.27\pm 0.30\pm 0.05$ \\
$\pi^+\pi^-$ & 50.0 $a_1^2\,|V_{ub}|^2$ & & \\
$\pi^+\varrho^- + \varrho^+\pi^-\!\!$ & 176.9 $a_1^2\,|V_{ub}|^2$ &
 & \\[0.1cm]
\hline
\end{tabular}
\end{center}
\end{table}

\begin{table}
\caption{\label{tab:B0newb}
Branching ratios (in \%) of class~II non-leptonic $\bar B^0$ decays in
the new model. We take $\theta=20^\circ$ for the $\eta$--$\eta'$ mixing
angle}
\vspace{0.4cm}
\begin{center}
\begin{tabular}{|l|cc|c|}
\hline
$\bar B^0$ Modes \rule{0cm}{0.4cm} & New Model &
 $a_1^{\rm eff}=0.98$ & Experimental \\
 & & $a_2^{\rm eff}=0.29$ & Average \\[0.1cm]
\hline
Class II \rule[-0.2cm]{0cm}{0.6cm} & & & \\
\hline
$\bar K^0 J/\psi$ \rule{0cm}{0.4cm} & 0.800 $a_2^2$ & 0.07 &
 $0.075\pm 0.021$ \\
$\bar K^0 \psi(\mbox{2S})$ & 0.326 $a_2^2$ & 0.03 & $<0.08$ \\
$\bar K^{*0} J/\psi$ & 2.518 $a_2^2$ & 0.21 & $0.153\pm 0.028$ \\
$\bar K^{*0} \psi(\mbox{2S})$ & 1.424 $a_2^2$ & 0.12 &
 $0.151\pm 0.091$ \\
$\pi^0 J/\psi$ & 0.018 $a_2^2$ & 0.002 & $<0.006$ \\
$\varrho^0 J/\psi$ & 0.050 $a_2^2$ & 0.004 & $<0.025$ \\
$\pi^0 D^0$ & 0.084 $a_2^2\,(f_D/200)^2$ & 0.007 & $<0.033$ \\
$\pi^0 D^{*0}$ & 0.116 $a_2^2\,(f_{D^*}/230)^2$ & 0.010 & $<0.055$ \\
$\varrho^0 D^0$ & 0.078 $a_2^2\,(f_D/200)^2$ & 0.007 & $<0.055$ \\
$\varrho^0 D^{*0}$ & 0.199 $a_2^2\,(f_{D^*}/230)^2$ & 0.017 & $<0.117$
\\
$\omega\,D^0$ & 0.081 $a_2^2\,(f_D/200)^2$ & 0.007 & $<0.057$ \\
$\omega\,D^{*0}$ & 0.203 $a_2^2\,(f_{D^*}/230)^2$ & 0.017 & $<0.120$ \\
$\eta\,D^0$ & 0.058 $a_2^2\,(f_D/200)^2$ & 0.005 & $<0.033$ \\
$\eta\,D^{*0}$ & 0.073 $a_2^2\,(f_{D^*}/230)^2$ & 0.006 & $<0.050$ \\
[0.1cm]
\hline
\end{tabular}
\end{center}
\end{table}

\begin{table}
\caption{\label{tab:Bminew}
Branching ratios (in \%) of non-leptonic $B^-$ decays in the new model}
\vspace{0.4cm}
\begin{center}
\begin{tabular}{|l|ccc|}
\hline
$B^-$ Modes \rule{0cm}{0.4cm} & New Model & $a_1^{\rm eff}=0.98$ &
 Experimental \\
 & & $a_2^{\rm eff}=0.29$ & Average \\[0.1cm]
\hline
Class I \rule[-0.2cm]{0cm}{0.6cm} & & & \\
\hline
$D^0 D^-$ \rule{0cm}{0.4cm} & 0.039 $a_1^2\,(f_D/200)^2$ &
 0.04 & \\
$D^0 D_s^-$ & 1.069 $a_1^2\,(f_{D_s}/240)^2$ & 1.03 &
 $1.36\pm 0.28\pm 0.33$ \\
$D^0 D^{*-}$ & 0.034 $a_1^2\,(f_{D^*}/230)^2$ & 0.03 & \\
$D^0 D^{*-}_s$ & 0.883 $a_1^2\,(f_{D_s^*}/275)^2$ & 0.85 &
 $0.94\pm 0.31\pm 0.23$ \\
$D^{*0} D^-$ & 0.025 $a_1^2\,(f_D/200)^2$ & 0.02 & \\
$D^{*0} D_s^-$ & 0.642 $a_1^2\,(f_{D_s}/240)^2$ & 0.62 &
 $1.18\pm 0.36\pm 0.29$ \\
$D^{*0} D^{*-}$ & 0.091 $a_1^2\,(f_{D^*}/230)^2$ & 0.09 & \\
$D^{*0} D^{*-}_s$ & 2.570 $a_1^2\,(f_{D_s^*}/275)^2$ & 2.47 &
 $2.70\pm 0.81\pm 0.66$ \\[0.1cm]
\hline
Class II \rule[-0.2cm]{0cm}{0.6cm} & & & \\
\hline
$K^- J/\psi$ \rule{0cm}{0.4cm} & 0.852 $a_2^2$ & 0.07 &
 $0.102\pm 0.014$ \\
$K^-\psi(\mbox{2S})$ & 0.347 $a_2^2$ & 0.03 & $0.070\pm 0.024$ \\
$K^{*-} J/\psi$ & 2.680 $a_2^2$ & 0.23 & $0.174\pm 0.047$ \\
$K^{*-} \psi(\mbox{2S})$ & 1.516 $a_2^2$ & 0.13 & $<0.30$ \\
$\pi^- J/\psi$ & 0.038 $a_2^2$ & 0.003 & $0.0057\pm 0.0026$ \\
$\varrho^- J/\psi$ & 0.107 $a_2^2$ & 0.009 & $<0.077$ \\[0.1cm]
\hline
Class III \rule[-0.2cm]{0cm}{0.6cm} & & & \\
\hline
$D^0\pi^-$ \rule{0cm}{0.4cm} & $0.338\left[a_1+0.729\,a_2\,
 (f_D/200)\right]^2$ & 0.48 & $0.50\pm 0.05\pm 0.02$ \\
$D^0\varrho^-$ & $0.828\left[a_1+0.450\,a_2\,(f_D/200)\right]^2$ &
 1.02 & $1.37\pm 0.18\pm 0.05$ \\
$D^0 a_1^-$ & $0.898\left[a_1+0.317\,a_2\,(f_D/200)\right]^2$ &
 1.03 & \\
$D^{*0}\pi^-$ & $0.315\left[a_1+0.886\,a_2\,(f_{D^*}/230)\right]^2$ &
 0.48 & $0.52\pm 0.08\pm 0.02$ \\
$D^{*0}\varrho^-$ & $0.926\,[a_1^2+0.456\,a_2^2\,(f_{D^*}/230)^2$ &
 1.26 & $1.51\pm 0.30\pm 0.02$ \\
 & $\phantom{0.926[} \mbox{}+ 1.292\,a_1 a_2\,(f_{D^*}/230)]$ & & \\
$D^{*0} a_1^-$ & $1.296\,[a_1^2+0.128\,a_2^2\,(f_{D^*}/230)^2$ &
 1.36 & $1.89\pm 0.53\pm 0.08$ \\
 & $\phantom{1.296[} \mbox{}+ 0.269\,a_1 a_2\,(f_{D^*}/230)]$ &
 & \\[0.1cm]
\hline
\end{tabular}
\end{center}
\end{table}

Let us now compare in detail the theoretical predictions with the data.
Due to the uncertainty in the values of the phenomenological parameters
$a_1^{\rm eff}$ and $a_2^{\rm eff}$, we first concentrate on ratios of
branching fractions in which these coefficients cancel. The comparison
of such predictions with data constitutes a test of the factorization
hypotheses and of the quality of our form factors. From the class~I
transitions listed in Tables~\ref{tab:B0decays} and \ref{tab:B0new}, we
obtain the predictions
\begin{eqnarray}
   R_1 &=& {{\rm B}(\bar B^0\to D^+\pi^-)\over
    {\rm B}(\bar B^0\to D^{*+}\pi^-)} \approx 1.04 \quad [1.07] \,,
    \nonumber\\
   R_2 &=& {{\rm B}(\bar B^0\to D^+\varrho^-)\over
    {\rm B}(\bar B^0\to D^{*+}\varrho^-)} \approx 0.88 \quad [0.89] \,.
\end{eqnarray}
Here and below we use the NRSX model as our nominal choice, and quote
results obtained with the new model of Section~\ref{sec:formfa} in
parentheses. In the heavy-quark limit both ratios become equal to
unity, because the spin symmetry relates $D$ and $D^*$. Only part of
the deviations from this limit is a form factor effect; the remainder
is of kinematic origin. The experimental results for these ratios are
$R_1=1.11\pm 0.23$ and $R_2=1.15\pm 0.34$, in agreement with our
predictions, although the errors are still large. We may also consider
the corresponding ratios for pairs of class~I transitions to final
states which differ only in their light meson. For instance, we find
\begin{eqnarray}
   R_3 &=& \left( \frac{f_\varrho}{f_\pi} \right)^2
    {{\rm B}(\bar B^0\to D^+\pi^-)\over
     {\rm B}(\bar B^0\to D^+\varrho^-)} \approx 1.03 \quad [1.06] \,,
    \nonumber\\
   R_4 &=& \left( \frac{f_\varrho}{f_\pi} \right)^2
    {{\rm B}(\bar B^0\to D^{*+}\pi^-)\over
     {\rm B}(\bar B^0\to D^{*+}\varrho^-)}\approx 0.88 \quad [0.88] \,,
\end{eqnarray}
which in the heavy-quark limit become equal unity, too. The
experimental values are $R_3=0.95\pm 0.24$ and $R_4=0.99\pm 0.25$,
again in agreement with our predictions.

Similar ratios can be taken for class~II amplitudes; however, since the
relevant matrix elements involve current matrix elements between a
heavy and a light meson, the theoretical predictions are considerably
more model dependent. Based on the results of our two models, we expect
\begin{eqnarray}
   R_5 &=& {{\rm B}(\bar B\to\bar K\,J/\psi)\over
    {\rm B}(\bar B\to\bar K^* J/\psi)} \approx 0.62 \quad [0.32] \,,
    \nonumber\\
   R_6 &=& {{\rm B}(\bar B\to\bar K\,\psi(\mbox{2S}))\over
    {\rm B}(\bar B\to\bar K^*\psi(\mbox{2S}))} \approx 0.54
    \quad [0.23] \,.
\end{eqnarray}
The strong model dependence does not allow for a test of the
factorization hypothesis in this case. The corresponding experimental
values are $R_5=0.58\pm 0.11$ and $R_6=0.44\pm 0.31$, where we have
averaged the results for $\bar B^0$ and $B^-$ decays (accounting for
the different lifetimes) when available. Although the experimental
errors are sizable, the data seem to prefer the NRSX model.

We now turn to the actual values of the phenomenological parameters
$a_1^{\rm eff}$ and $a_2^{\rm eff}$. From the theoretical point of
view, the cleanest determination of $a_1^{\rm eff}$ is from the class
of decays $\bar B^0\to D^{(*)+} h^-$, where $h^-$ is a light meson
($h=\pi$, $\varrho$ or $a_1$). The relevant $\bar B^0\to D^{(*)+}$
transition form factors are known from the analysis of semileptonic $B$
decays using the heavy-quark effective theory, and the decay constants
of the light mesons are experimentally known with good accuracy. Also,
these transitions have a similar decay kinematics, so that we may
expect that they are characterized by the same value of $a_1^{\rm
eff}$. Performing a fit to the experimental data, we
obtain\footnote{Since we are neglecting final-state interactions, the
parameters $a_i^{\rm eff}$ are real numbers, and by convention we take
$a_1^{\rm eff}$ to be positive.}
\begin{equation}
  a_1^{\rm eff}|_{Dh} = 1.08\pm 0.04 \quad [0.98\pm 0.04] \,.
\label{a1first}
\end{equation}
The coefficient $a_1^{\rm eff}$ can also be determined from the decays
$\bar B\to D^{(*)} D_s^{(*)-}$, which are characterized by a quite
different decay kinematics. In principle, it would be interesting to
investigate whether the resulting value is different in the two cases,
i.e.\ whether there is an observable process dependence of the
phenomenological parameter. In practise, this cannot be done because of
the large uncertainty in the values of the decay constants of charm
mesons. From a fit to the data, we find
\begin{equation}
  a_1^{\rm eff}|_{D D_s} = 1.10\pm 0.07\pm 0.17 \quad
  [1.05\pm 0.07\pm 0.16] \,,
\label{a1sec}
\end{equation}
where the second error accounts for the uncertainty in $f_{D_s^{(*)}}$.
In both cases, (\ref{a1first}) and (\ref{a1sec}), the data support the
theoretical expectation that $a_1^{\rm eff}$ is close to unity [see
(\ref{a1a2appr})].

A value for the parameter $|a_2^{\rm eff}|$ (but not the relative sign
between $a_2^{\rm eff}$ and $a_1^{\rm eff}$) can be obtained from the
class~II decays $\bar B\to\bar K^{(*)} J/\psi$ und $\bar B\to\bar
K^{(*)}\psi(\mbox{2S})$. From a fit to the six measured branching
ratios, we extract
\begin{equation}
   |a_2^{\rm eff}|_{K\psi} = 0.21\pm 0.01 \quad [0.29\pm 0.01] \,.
\label{a2first}
\end{equation}
This result is more strongly dependent on the form-factor model used,
which is not surprising given that class~II decays involve
heavy-to-light transition matrix elements. As we have seen above, the
NRSX model may be the more trustable one; still, we believe the
difference between the two results provides a realistic estimate of the
theoretical uncertainty.

A determination of $a_2^{\rm eff}$ from decays with a rather different
kinematics is possible by considering the class~III transitions $B^-\to
D^{(*)0} h^-$ with $h=\pi$ or $\varrho$. Moreover, because of the
interference of $a_1$ and $a_2$ amplitudes, these decays are sensitive
to the relative sign of the QCD coefficients. From the theoretical
point of view, it is of advantage to normalize the branching ratios to
those of the corresponding $\bar B^0$ decays, which are class~I
transitions. The theoretical predictions for these ratios are of the
form
\begin{equation}
   \frac{{\rm B}(B^-\to D^{(*)0} h^-)}
    {{\rm B}(\bar B^0\to D^{(*)+} h^-)}
   = \frac{\tau(B^-)}{\tau(\bar B^0)} \left[ 1 + 2 x_1\,
   \frac{a_2^{\rm eff}}{a_1^{\rm eff}} + x_2^2 \left(
   \frac{a_2^{\rm eff}}{a_1^{\rm eff}} \right)^2 \right] \,,
\end{equation}
where $x_1$ and $x_2$ are process-dependent parameters depending on
the ratio of some hadronic form factors and decay constants ($x_1=x_2$
except for the decay $B^-\to D^{*0}\varrho^-$). For the ratios of
branching fractions on the left-hand side, we use recent CLEO data
reported in Ref.~86,
which are more accurate than the
previous world averages presented in
Tables~\ref{tab:B0decays}--\ref{tab:Bminew}. Performing a fit to the
data, we extract the ratio $a_2^{\rm eff}/a_1^{\rm eff}$ for each
channel. The results are collected in Table~\ref{tab:vgl}, where the
second error results from the uncertainty in the lifetime
ratio~\cite{Rich} $\tau(B^-)/\tau(\bar B^0)=1.06\pm 0.04$. Taking the
average, and using (\ref{a1first}), we find
\begin{eqnarray}
   \left. \frac{a_2^{\rm eff}}{a_1^{\rm eff}}\right|_{Dh}
   &=& 0.21\pm 0.05 \quad [0.31\pm 0.08] \,, \nonumber\\
   a_2^{\rm eff}|_{Dh}
   &=& 0.23\pm 0.05 \quad [0.30\pm 0.08] \,.
\label{a2sec}
\end{eqnarray}
The value of $a_2^{\rm eff}$ is in good agreement with that obtained in
(\ref{a2first}). Given the different decay kinematics in the two
processes, this observation is quite remarkable.

\begin{table}
\caption{\label{tab:vgl}
Ratios of non-leptonic decay rates of charged and neutral $B$
mesons~\protect\cite{Rodri}, and the corresponding values for $a_2^{\rm
eff}/a_1^{\rm eff}$}
\vspace{0.4cm}
\begin{center}
\begin{tabular}{|l|c|c|}\hline
Experimental Ratios \rule[-0.2cm]{0cm}{0.6cm} &
 Predictions for $x_i$ & $a_2^{\rm eff}/a_1^{\rm eff}$ \\
\hline
$\displaystyle{{\rm B}(B^-\to D^0\pi^-)
 \over{\rm B}(\bar B^0\to D^+\pi^-)} = 1.73\pm 0.25$
 \rule{0cm}{0.6cm} &
 $1.127\quad[0.729]$ & $0.24\pm 0.08\pm 0.02$ \\
 & & $[0.38\pm 0.13\pm 0.03]$ \\[0.1cm]
$\displaystyle{{\rm B}(B^-\to D^0\varrho^-)
 \over{\rm B}(\bar B^0\to D^+\varrho^-)} = 1.19\pm 0.24$ &
 $0.587\quad[0.450]$ & $0.10\pm 0.18\pm 0.03$ \\
 & & $[0.13\pm 0.24\pm 0.04]$ \\[0.1cm]
$\displaystyle{{\rm B}(B^-\to D^{*0}\pi^-)
 \over{\rm B}(\bar B^0\to D^{*+}\pi^-)} = 1.64\pm 0.28$ &
 $1.361\quad[0.886]$ & $0.18\pm 0.08\pm 0.02$ \\
 & & $[0.27\pm 0.12\pm 0.03]$ \\[0.1cm]
$\displaystyle{{\rm B}(B^-\to D^{*0}\varrho^-)
 \over{\rm B}(\bar B^0\to D^{*+}\varrho^-)} = 1.71\pm 0.36$ &
 $x_1=0.759~[0.646]$ & $0.35\pm 0.17\pm 0.03$ \\
 & $x_2=0.813~[0.675]$ & $[0.41\pm 0.20\pm 0.04]$ \\[0.1cm]
\hline
\end{tabular}
\end{center}
\end{table}

The magnitude and, in particular, the positive sign of $a_2^{\rm eff}$
are of great importance for the theoretical interpretation of our
results. We find that in non-leptonic $B$ decays the two parameters
$a_1^{\rm eff}$ and $a_2^{\rm eff}$ have the same sign, meaning that
the corresponding amplitudes interfere constructively. This finding is
in stark contrast to the situation encountered in $D$-meson decays,
where a similar analysis yields~\cite{bsw87,firstedi}
\begin{equation}
   a_1^{\rm eff}|_{\rm charm} = 1.10\pm 0.05 \,, \qquad
   a_2^{\rm eff}|_{\rm charm} = -0.49\pm 0.04 \,,
\end{equation}
indicating a strong destructive interference. Since most $D$ decays are
(quasi) two-body transitions, this effect is responsible for the
observed lifetime difference between $D^+$ and $D^0$ mesons~\cite{PDG}:
$\tau(D^+)/\tau(D^0)=2.55\pm 0.04$. In $B$ decays, on the other hand,
the majority of transitions proceeds into multi-body final states, and
moreover there are many $B^-$ decays (such involving two charm quarks
in the final state) where no interference can occur. The relevant scale
for multi-body decay modes may be significantly lower than $m_b$,
leading to destructive interference (see Figure~\ref{Fig:a2a1}).
Therefore, the observed constructive interference in the two-body modes
is not in conflict with the fact that $\tau(B^-)>\tau(\bar B^0)$.

The values for $a_2^{\rm eff}$ extracted from $\bar B\to\bar
K^{(*)}\psi$ and $\bar B\to D^{(*)}h$ decays in (\ref{a2first}) and
(\ref{a2sec}) indicate that non-factorizable contributions (at the
scale $\mu=m_b$) are small in these processes. Using $a_2^{\rm
eff}|_{K\psi}=c_2(m_b)+\zeta_{K\psi} c_1(m_b)=0.21\pm 0.05$ and
$a_2^{\rm eff}|_{Dh}=c_2(m_b)+\zeta_{Dh} c_1(m_b)=0.23\pm 0.05$ with
conservative errors, and combining these with the values of the Wilson
coefficients given in Table~\ref{tab:c1c2}, we find
\begin{eqnarray}
   \zeta_{K\psi} &=& 0.44\pm 0.05 \,, \qquad
    \varepsilon_8^{(BK,\psi)}(m_b) = 0.11\pm 0.05 \,, \nonumber\\
   \zeta_{Dh} &=& 0.46\pm 0.05 \,, \qquad
    \varepsilon_8^{(BD,h)}(m_b) = 0.13\pm 0.05 \,.
\end{eqnarray}
Hence, within errors there is no experimental evidence for a process
dependence of the value of $\zeta$, in accordance with our expectation
stated in (\ref{delzeta}).

\section{Tests of Factorization and Extraction of Decay Constants}
\label{sec:tests}

Based on the factorization hypothesis, we have made in the previous
section a number of predictions for the ratios of hadronic decay rates.
Within the current experimental uncertainties, these predictions agree
well with the available data. In this section, we shall discuss a
particularly clean method to test the factorization hypothesis. As
suggested by Bjorken~\cite{Bj89}, we make use of the close relationship
between semileptonic and factorized hadronic amplitudes by dividing the
non-leptonic decay rates by the corresponding differential semileptonic
decay rates evaluated at the same $q^2$. This method provides a direct
test of the factorization hypothesis; moreover, it may be used to
determine some interesting decay constants~\cite{bo90,rosner}. Assuming
that factorization holds, we have
\begin{equation}\label{RP}
   R_M^{(*)} = {{\rm B}(\bar B^0\to D^{(*)+} M^-)\over
   {\rm d}{\rm B}(\bar B^0\to D^{(*)+}\ell^-\bar\nu)/{\rm d}q^2
   \Big|_{q^2=m_M^2}} = 6\pi^2 f_M^2\,|a_1^{\rm eff}|^2\,|V_{ij}|^2\,
   X_M^{(*)} \,,
\end{equation}
with $a_1^{\rm eff}\approx 1$. Here $f_M$ is the decay constant of the
meson $M$, and $V_{ij}$ is the appropriate CKM matrix element
(depending on the flavour quantum numbers of the meson $M$). To
determine this ratio experimentally, one needs the values of the
differential semileptonic branching ratio at various values of $q^2$.
They have been determined for $\bar B\to D^*\ell\,\bar\nu$ decays in
Ref.~81,
using a fit to experimental data. We collected
their results in Table~\ref{tab:dGdq2}.

\begin{table}
\caption{\label{tab:dGdq2}
Values of the differential semileptonic branching ratio used in the
factorization tests, as compiled in Ref.~81
}
\vspace{0.4cm}
\begin{center}
\begin{tabular}{|lc|}\hline
\rule[-0.4cm]{0cm}{1.0cm}
$q^2$ & $\displaystyle{{\rm d}{\rm B}(\bar B\to D^*\ell\,\bar\nu)
 \over {\rm d}q^2}~[10^{-2}~\mbox{GeV}^{-2}]$ \\
\hline
$m_\pi^2$ \rule{0cm}{0.4cm} & $0.237\pm 0.026$ \\[0.1cm]
$m_\varrho^2$ & $0.250\pm 0.030$ \\[0.1cm]
$m_{a_1}^2$   & $0.335\pm 0.033$ \\[0.1cm]
$m_{D_s}^2$   & $0.483\pm 0.033$ \\[0.1cm]
$m_{D_s^*}^2$ & $0.507\pm 0.035$ \\[0.1cm]
\hline
\end{tabular}
\end{center}
\end{table}

Neglecting the lepton mass, we obtain for a pseudoscalar meson $P$:
\begin{eqnarray}\label{XP}
   X_P &=& {(m_B^2-m_D^2)^2\over
    [m_B^2-(m_D+m_P)^2]\,[m_B^2-(m_D-m_P)^2]}\,
    \left| {F_0(m_P^2)\over F_1(m_P^2)} \right|^2 \,,\nonumber\\
   X_P^* &=& \left[ m_B^2-(m_{D^*}+m_P)^2 \right]\,
    \left[ m_B^2-(m_{D^*}-m_P)^2 \right] \nonumber\\
   &&\times
    {|A_0(m_P^2)|^2\over m_P^2\,\sum_{i=0,\pm}\,|H_i(m_P^2)|^2} \,.
\end{eqnarray}
The helicity amplitudes $H_0(q^2)$ and $H_{\pm}(q^2)$ are defined in
the Appendix. For the special case that the pseudoscalar meson is a
pion, it is an excellent approximation to expand the quantities
$X_{\pi}$ and $X_{\pi}^*$ in powers of $m_\pi^2/m_B^2$,
yielding~\cite{firstedi}
\begin{eqnarray}
   X_\pi &\simeq& 1 + {4m_\pi^2 m_B m_D\over (m_B^2-m_D^2)^2}
    \approx 1.001 \,, \nonumber\\
   X_\pi^* &\simeq& 1+ {4 m_{\pi}^2 m_B m_{D^*}\over
    (m_B^2-m_{D^*}^2)^2} - {4 m_{\pi}^2\over (m_B-m_{D^*})^2}
    \approx 0.994 \,.
\end{eqnarray}
Neglecting the tiny deviation from unity, we obtain
\begin{equation}\label{76}
   R_\pi = R_\pi^* = 6\pi^2 f_\pi^2\,|a_1^{\rm eff}|^2\,|V_{ud}|^2
   \approx |a_1^{\rm eff}|^2 \times 0.96~\mbox{GeV}^2 \,.
\end{equation}
This prediction may be compared with the experimental value
$R_\pi^*=1.18\pm 0.22$~GeV$^2$ obtained by combining the $\bar B^0\to
D^{*+}\pi^-$ branching ratio from Table~\ref{tab:B0decays} with the
value for the differential semileptonic branching ratio at
$q^2=m_\pi^2$ given in Table~\ref{tab:dGdq2}. This yields $a_1^{\rm
eff}=1.11\pm 0.10$, in good agreement with the expectation based on
factorization. An even cleaner test of factorization is obtained when
(\ref{RP}) is evaluated for a vector or pseudovector meson, in which
case one has exactly~\cite{rie3}
\begin{equation}
   X_V = X_V^* = 1 \,.
\end{equation}
Since the lepton pair created by the $(V-A)$ current carries spin one,
its production is kinematically equivalent to that of a (pseudo-)
vector particle with four-momentum $q_\mu$. For a $\varrho$ meson in
the final state, for instance, we thus obtain
\begin{equation}\label{77}
   R_\varrho = R_\varrho^* = 6\pi^2 f_\varrho^2\,|a_1^{\rm eff}|^2\,
   |V_{ud}|^2 \approx |a_1^{\rm eff}|^2 \times 2.48~\mbox{GeV}^2 \,,
\end{equation}
to be compared with the experimental value $R_\varrho^*=2.92\pm
0.71$~GeV$^2$. This gives $a_1^{\rm eff}=1.09\pm 0.13$, again in good
agreement with the expectation based on factorization. In principle,
eqs.~(\ref{76}) and (\ref{77}) offer the possibility for four
independent determinations of the QCD parameter $a_1^{\rm eff}$. Good
agreement among the extracted values supports the validity of the
factorization approximation in $B$ decays. Already at the present level
of accuracy, it shows that there is little room for final-state
interactions affecting the magnitude of the considered decay
amplitudes.

{}From the kinematic argument about the equivalence of the lepton pair
in the semileptonic decay and the spin-1 meson in the hadronic decay,
it follows that (\ref{RP}) is valid separately for longitudinal and
transverse polarization of the $D^*$ meson in the final state. Thus,
the polarization of the $D^*$ meson produced in the non-leptonic decay
$\bar B^0\to D^{*+} V^-$ should be equal to the polarization in the
corresponding semileptonic decay $\bar B\to D^*\ell\,\bar\nu$ at
$q^2=m_V^2$. However, in order to turn this prediction into a test of
the factorization hypothesis one would have to determine the
polarization of the $D^*$ meson with high precision. This is so because
in the semileptonic as well as in the non-leptonic case the $D^*$
polarization at the points $q^2=0$ and $q^2=q^2_{\rm max}$ is
determined by kinematics alone to be 100\% longitudinal and $1/3$
longitudinal, respectively. This shows that for a stringent test of the
factorization hypothesis at small $q^2$ one must determine the
transverse polarization contribution with a small relative error.
Especially for the semileptonic decays, a precision measurement of the
$q^2$ dependence of the polarization appears to be a complicated task,
however. Still, we can make a rather precise prediction for this
quantity using heavy-quark symmetry. In the heavy-quark limit, the
ratio of transverse to longitudinal polarization at some fixed $q^2$ is
simply given by
\begin{equation}\label{pol}
   {\Gamma_{\rm T}\over\Gamma_{\rm L}}
   = {4 q^2(m_B^2+m_{D^*}^2-q^2)\over (m_B-m_{D^*})^2
    [(m_B+m_{D^*})^2-q^2]} \,.
\end{equation}
Including the leading symmetry-breaking corrections to this
result~\cite{qcd}$^-$\cite{Volker}, one obtains the numbers shown in
Table~\ref{tab:dGdG}. For the polarization of the $D^*$ meson in the
decay $\bar B^0\to D^{*+}\varrho^-$, the CLEO Collaboration
finds~\cite{honreview} $\Gamma_{\rm T}/\Gamma_{\rm tot}=(7\pm 5\pm
5)\%$, in agreement with our prediction of 12\% transverse polarization
for the semileptonic decay at $q^2=m_\varrho^2$. However, in order for
this test to be sensitive to deviations from factorization, the
experimental uncertainty will have to be reduced substantially . The
situation may be more favourable in the decay $\bar B^0\to D^{*+}
D_s^*$ with predicted 48\% of transverse polarization, hopefully
allowing for a measurement with smaller relative uncertainties.

\begin{table}
\caption{\label{tab:dGdG}
Theoretical predictions for the ratio $\Gamma_{\rm T}/\Gamma_{\rm tot}$
at fixed $q^2$, where $\Gamma_{\rm tot}=\Gamma_{\rm T}+\Gamma_{\rm L}$}
\vspace{0.4cm}
\begin{center}
\begin{tabular}{|c|ccccc|}\hline
$q^2$ & \rule[-0.2cm]{0cm}{0.6cm} 0 & $m_\varrho^2$ & $m_{a_1}^2$ &
 $m_{D_s^*}^2$ & $q_{\rm max}^2$ \\
\hline
$\Gamma_{\rm T}/\Gamma_{\rm tot}$ & \rule[-0.2cm]{0cm}{0.6cm} 0 &
 $12\pm 1$ & $26\pm 2$ & $48\pm 1$ & 2/3 \\
\hline
\end{tabular}
\end{center}
\end{table}

We shall now discuss an alternative use of (\ref{RP}). Assuming the
validity of the factorization hypothesis with a fixed value for
$a_1^{\rm eff}$, one may employ this relation for the determination of
unknown decay constants~\cite{bo90,rosner}. In particular, we can use
it to determine the decay constants of the $D_s$ and $D_s^*$ mesons. As
explained above, in the latter case we simply have
$X_{D_s^*}=X^*_{D_s^*}=1$. But also in the case of $D_s$ mesons we
obtain essentially model-independent predictions. Defining the mass
ratio $x=\frac 12(m_{D^{(*)}}+m_{D_s})/m_B$, we find in the heavy-quark
limit~\cite{firstedi} ($x\approx 0.36$ and $x\approx 0.38$,
respectively)
\begin{eqnarray}
    X_{D_s} &=& {1-3x^2+2x^3\over 1-3x^2-2x^3}
     \approx 1.36 \,,\nonumber\\
    X_{D_s}^* &=& {1-x-2x^2\over 1-x+2x^2} \approx 0.37 \,.
\end{eqnarray}
These values are close to the predictions of the NRSX model, which are
\begin{equation}\label{numerical}
   X_{D_s} \approx  1.33 \,,\qquad X_{D_s}^* \approx 0.39 \,.
\end{equation}
Using these values (allowing for a theoretical error of $\pm 0.03$)
together with $|a_1^{\rm eff}|=1.08\pm 0.04$, we obtain the theoretical
predictions:
\begin{eqnarray}
   R_{D_s} &=& (87.2\pm 6.7)\,f_{D_s}^2 \,, \nonumber\\
   R^*_{D_s} &=& (25.6\pm 2.7)\,f_{D_s}^2 \,, \nonumber\\
   R_{D_s^*} &=& R^*_{D_s^*} = (65.6\pm 4.8)\,f_{D_s^*}^2 \,.
\label{Rthe}
\end{eqnarray}
By averaging the experimental data on $\bar B^0$ and $B^-$ decays into
two charm mesons (taking into account the differences in the $B$-meson
lifetimes), we obtain
\begin{eqnarray}
   {\rm B}(\bar B\to D D_s^-) &=& (0.95\pm 0.24)\% \,,
    \nonumber\\
   {\rm B}(\bar B\to D D_s^{*-}) &=& (1.00\pm 0.30)\% \,,
    \nonumber\\
   {\rm B}(\bar B\to D^* D_s^-) &=& (1.03\pm 0.27)\% \,,
    \nonumber\\
   {\rm B}(\bar B\to D^* D_s^{*-}) &=& (2.26\pm 0.60)\% \,.
\label{expav}
\end{eqnarray}
Using the last two averaged decay rates gives the experimental ratios
\begin{equation}
   R^*_{D_s} = (2.13\pm 0.58)~\mbox{GeV}^2 \,, \qquad
   R^*_{D_s^*} = (4.46\pm 1.22)~\mbox{GeV}^2 \,.
\end{equation}
Comparing this with the theoretical predictions in (\ref{Rthe}), we
find
\begin{equation}\label{fds37}
   f_{D_s} = (288\pm 42)~{\rm MeV} \,, \qquad
   f_{D_s^*} = (261\pm 36)~{\rm MeV} \,.
\end{equation}

More precise determinations of the decay constants are possible if,
instead of using (\ref{RP}), we consider ratios of non-leptonic decay
rates, comparing processes involving $D_s$ and $D_s^*$ mesons with
those involving the light mesons $\pi$ and $\varrho$. These processes
involve a similar kinematics, so that the ratios of the corresponding
decay rates are sensitive to the same form factors, however evaluated
at different $q^2$ values. This method has the advantage that in the
ratios the phenomenological parameter $a_1^{\rm eff}$ cancels;
similarly, we may hope that some of the experimental systematic errors
cancel. Using the NRSX model, we find
\begin{eqnarray}
   {{\rm B}(\bar B^0\to D^+ D_s^-)\over
    {\rm B}(\bar B^0\to D^+\pi^-)}
   &=& 1.01\,\left( {f_{D_s}\over f_\pi} \right)^2 \,, \nonumber\\
   {{\rm B}(\bar B^0\to D^{*+} D_s^-)\over
    {\rm B}(\bar B^0\to D^{*+}\pi^-)}
   &=& 0.72\,\left( {f_{D_s}\over f_\pi} \right)^2 \,, \nonumber\\
   {{\rm B}(\bar B^0\to D^+ D_s^{*-})\over
    {\rm B}(\bar B^0\to D^+\varrho^-)}
   &=& 0.74\,\left( {f_{D_s^*}\over f_\varrho} \right)^2 \,,
\nonumber\\
   {{\rm B}(\bar B^0\to D^{*+} D_s^{*-})\over
    {\rm B}(\bar B^0\to D^{*+}\varrho^-)}
   &=& 1.68\,\left( {f_{D_s^*}\over f_\varrho} \right)^2 \,.
\end{eqnarray}
Combining these predictions with the average experimental branching
ratios in (\ref{expav}), we find the rather accurate values
\begin{equation}
   f_{D_s} = (234\pm 25)~{\rm MeV} \,, \qquad
   f_{D_s^*} = (271\pm 33)~{\rm MeV} \,.
\end{equation}
The result for $f_{D_s}$ is in excellent agreement with the value
$f_{D_s}=241\pm 37$~MeV in (\ref{fDs}), extracted from leptonic decays
of $D_s$ mesons. The ratio $f_{D_s^*}/f_{D_s}=1.16\pm 0.19$, which
cannot be determined from leptonic decays, is in good agreement with
theoretical expectations~\cite{matthias,abada}.

Along these lines, there are numerous other possibilities for
extracting information on decay constants and current matrix elements.
In particular, the decay constants of P-wave particles like $a_0$,
$a_1$, $K_0^*$ and $K_1$ are of interest due to their sensitivity to
relativistic quark motion, which allows for a test of hadron models.
For instance, one may use the ratio
\begin{equation}
   {{\rm B}(\bar B^0\to D^{*+} a_1^-)\over
    {\rm B}(\bar B^0\to D^{*+}\varrho^-)}
   \approx 1.18\,\left( {f_{a_1}\over f_\varrho} \right)^2
\end{equation}
to determine the pseudovector meson decay constant $f_{a_1}$, which is
defined by $\langle\,0\,|\,\bar q\gamma_\mu\gamma_5 q\,|a_1\rangle =
\epsilon_\mu m_{a_1} f_{a_1}$. From the experimental results, we obtain
\begin{equation}
   f_{a_1} = (1.22\pm 0.19)\,f_\varrho = (256\pm 40)~{\rm MeV} \,,
\end{equation}
which agrees with the large value derived from $\tau$ decays shown in
Table~\ref{tab:fM}. The fact that the decay constant of a P-wave meson
is of the same order of magnitude as those of the corresponding S-wave
mesons is quite remarkable. It shows that mesons containing light
quarks are highly relativistic systems; for non-relativistic
constituent quarks one would expect the wave function of a state with
non-zero orbital momentum to vanish at the origin, leading to a
vanishing decay constant.

\boldmath
\section{$B$ Decays to Baryons}
\unboldmath
\label{sec:baryons}

\begin{figure}
\epsfysize=3.5cm
\centerline{\epsffile{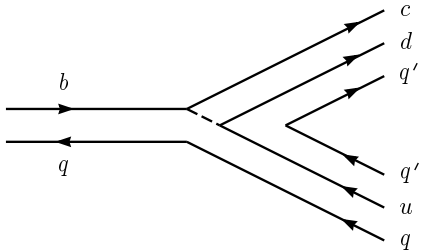}}
\caption{\label{Fig:baryons}
Flavour flow diagram for a $B$ decay into a baryon-antibaryon pair.}
\end{figure}

We end our discussion of non-leptonic processes with a remark on the
particularly interesting case of $B$ decays into baryon-antibaryon
pairs. For these transitions it is probably not sufficient to apply
factorization in the way described so far. The relevant flavour flow
diagram (for a $b\to c$ transition) is shown in
Figure~\ref{Fig:baryons}. In contrast with the situation realized in
$B$ decays into mesons, the $c$ and $d$ quarks produced in the weak
interaction now end up in the same hadron. It is, therefore,
appropriate to rewrite the effective weak Hamiltonian in a different
form using charge-conjugate fields and a Fierz reordering~\cite{St87}:
\begin{eqnarray}\label{Hdiq}
   H_{\rm eff} &=& {G_F\over\sqrt{2}}\,V_{cb}\,\left[
    c_-(\mu)\,(c d')_{3^*}^\dagger (u b)_{3^*}
    + c_+(\mu)\,(c d')_6^\dagger (u b)_{6} \right] + \mbox{h.c.}
    \nonumber\\
   &&\mbox{}+ \mbox{penguin operators} \,,
\end{eqnarray}
where $(c d')_{3^*}=\varepsilon _{kij}\,\bar c_i^c\,(1-\gamma_5)\,
d'_j$ is an $(S-P)$ colour-antitriplet diquark current. The
colour-sextet current $(c d')_6$ is given by a similar expression. The
appearance of the coefficients $c_-(\mu)$ and $c_+(\mu)$ shows that
perturbative QCD interactions affect the colour-antitriplet and sextet
channels differently: the QCD force is attractive $(c_->1)$ in the
antitriplet and repulsive $(c_+<1)$ in the sextet channel.

{}From the very existence of baryons it is evident that, besides the
perturbative enhancement, there must also be a strong long-distance
binding force between two quarks in a colour-antitriplet state. This
can be taken into account by introducing diquark fields and their
couplings to the local currents occurring in (\ref{Hdiq}). QCD sum
rules predict that these diquark couplings are large~\cite{Ya89}. This
result has been confirmed by the magnitude of $\Delta S=1$ transitions,
where diquark effects play a dominant role~\cite{diquarks}. Thus, we
expect $B$ decays to baryon-antibaryon pairs to proceed predominantly
via the formation of a diquark-antidiquark state, followed by the
creation of a quark-antiquark pair in the colour field of these
diquarks. This decay mechanism implies interesting selection rules,
which allow for a number of predictions~\cite{iso,Ball}. For instance,
the charm baryon in baryonic $b\to c$ transitions is built from $(c d)$
spin-zero states. Thus, in $\bar B$ decays one should not find a $(c u
u)$ nor a $(c u s)$ nor a $(c s s)$ baryon, as long as final-state
interactions may be neglected. In $b\to u$ transitions one naturally
obtains a $\Delta I=\frac{1}{2}$ selection rule, because a spin-zero
$(u d)$ diquark necessarily has isospin zero. No $\Delta$ resonances
should therefore be produced in these decays, but there is no such
suppression of $\bar \Delta$ production. A comparison of these
predictions with experimental data, once these will become available,
should provide information about the creation process of
quark-antiquark pairs inside hadrons. For an estimate of the expected
branching ratios, the reader is referred to Ref.~114.

\section{Summary}
\label{sec:sum}

We have presented an overview of the theory and phenomenology of
exclusive hadronic decays of $B$ mesons, concentrating on two-body
modes. Such decays are strongly influenced by the long-range QCD colour
forces. Theoretically, their description involves hadronic matrix
elements of local four-quark operators, which are notoriously difficult
to calculate. The factorization approximation is used to relate these
matrix elements to products of current matrix elements. Conventionally,
the factorized decay amplitudes depend on two phenomenological
parameters $a_1$ and $a_2$, which are connected with the Wilson
coefficients $c_i(\mu)$ appearing in the effective weak Hamiltonian. We
have shown that (except for decays into two vector mesons) this
approach can be generalized in a natural way to include the dominant
non-factorizable contributions to the decay amplitudes. In the
generalized factorization scheme, the effective parameters $a_1^{\rm
eff}$ and $a_2^{\rm eff}$ become process-dependent. However, using the
large-$N_c$ counting rules of QCD we have argued that in energetic
two-body decays of $B$ mesons $a_1^{\rm eff}\approx 1$ and $a_2^{\rm
eff}\approx c_2(m_b)+\zeta\,c_1(m_b)$, where $\zeta=O(1/N_c)$ is a
dynamical parameter. Moreover, we have shown that the process
dependence of $\zeta$ is likely to be very mild, so that it can be
taken to be a constant for a wide class of two-body decays. These
theoretical expectations are supported by the data. From a fit to the
world average branching ratios of two-body decay modes, we obtain
$a_1^{\rm eff}\approx 1.08$ and $a_2^{\rm eff}\approx 0.21$,
corresponding to $\zeta\approx 0.45$. There is no evidence for a
process dependence of these parameters; in particular, the values
obtained for $a_2^{\rm eff}$ from the decays $\bar B\to\bar
K^{(*)}\psi$ and $B^-\to D^{(*)0} h^-$, where $h=\pi$ or $\varrho$, are
in good agreement with each other.

The obvious interpretation of the fact that the ratio $a_2^{\rm
eff}/a_1^{\rm eff}$ is positive and the value of the parameter $\zeta$
close to the ``naive'' factorization prediction $\zeta=1/3$ is that, in
energetic two-body decays of $B$ mesons, a fast moving colour-singlet
quark pair interacts little with soft gluons. Hence, factorization
works at a high scale of order $m_b$. The situation is different from
that encountered in the much less energetic decays of $D$ mesons, where
one finds a negative value of $a_2^{\rm eff}/a_1^{\rm eff}$,
corresponding to $\zeta\approx 0$. Since the energy release in charm
decays is much less, even soft gluons can rearrange the quarks, and the
effective factorization scale is lower. According to the relation
between the factorization scale and the ratio of the phenomenological
parameters exhibited in Figure~\ref{Fig:a2a1}, this leads to a negative
value of $a_2^{\rm eff}/a_1^{\rm eff}$ and thus a smaller value of
$\zeta$.

The most important ingredient of factorized decay amplitudes are the
hadronic form factors parametrizing the hadronic matrix elements of
quark currents. In the case of the heavy-to-heavy transitions $\bar
B\to D$ and $\bar B\to D^*$, heavy-quark symmetry implies simple
relations between the various form factors. Incorporating the leading
symmetry-breaking corrections using the heavy-quark effective theory,
it has become possible to extract all $\bar B\to D^{(*)}$ form factors
from semileptonic decay data with good precision. The use of
heavy-quark symmetry constitutes a significant improvement over earlier
estimates of non-leptonic amplitudes, which were based on model
calculations of the relevant form factors. As a consequence, for all
class~I decays considered in this article the factorized decay
amplitudes can be predicted without any model assumptions.

There has not been similar progress in the calculation of current
matrix elements between heavy and light mesons. For these transitions
we must still rely on phenomenological models. Consequently, the
theoretical predictions for class~II decay amplitudes involve larger
theoretical uncertainties. In order to get an idea about the amount of
model dependence, we have considered two different quark models and
compared their results. We find that it is possible to determine the
parameter $a_2^{\rm eff}$ with a theoretical accuracy of about $25\%$.
In the context of each model, a large set of class~II branching ratios
can be reproduced within the experimental errors using a fixed value of
$a_2^{\rm eff}$.

We have discussed various tests of the (generalized) factorization
hypothesis by considering ratios of decay rates, and by comparing
non-leptonic decay rates with semileptonic rates evaluated at the same
value of $q^2$. Within the present experimental uncertainties, there
are no indications for any deviations from the factorization scheme in
which $a_1^{\rm eff}$ and $a_2^{\rm eff}$ are treated as
process-independent hadronic parameters. Accepting that this scheme
provides a useful phenomenological concept, exclusive two-body decays
of $B$ mesons offer a unique opportunity to measure the decay constants
of some light or charm mesons, such as the $a_1$, $D_s$ and $D_s^*$. In
particular, on the basis of the experimental data available today, we
find $f_{D_s}=(234\pm 25)$~MeV and $f_{D_s^*}=(271\pm 33)$~MeV. The
result for $f_{D_s}$ is in excellent agreement with the value extracted
from leptonic decays of $D_s$ mesons. The ratio
$f_{D_s^*}/f_{D_s}=1.16\pm 0.19$ is in good agreement with theoretical
expectations.

\newpage
\section*{Acknowledgments}
We wish to thank K.~Honscheid for providing us with the most up-to-date
experimental results on $B$ decay branching ratios. Part of the work
presented here has been done in a most enjoyable collaboration with
V.~Rieckert and Q.P.~Xu, which is gratefully acknowledged. This work
has been supported in part by the Bundesministerium f\"ur Forschung und
Technologie, Bonn, Germany.

\setcounter{equation}{0}
\renewcommand{\theequation}{A.\arabic{equation}}
\section*{Appendix}
\label{sec:app}

We collect our definitions for the weak decay form factors, which
parametrize the hadronic matrix elements matrix elements of
flavour-changing vector and axial currents between meson
states. For the transition between two pseudoscalar mesons,
$P_1(p)\to P_2(p')$, we define
\begin{equation}\label{ffdef1}
   \langle P_2(p')|\,V_\mu |P_1(p)\rangle
   = \left( (p+p')_\mu - {m_1^2-m_2^2\over q^2}\,q_\mu \right)
   F_1(q^2) + {m_1^2-m_2^2\over q^2}\,q_\mu\,F_0(q^2) \,,
\end{equation}
where $q_\mu=(p-p')_\mu$ is the momentum transfer. For the transition
of a pseudoscalar into a vector meson, $P_1(p)\to V_2(\epsilon,p')$,
we define
\begin{eqnarray}\label{ffdef2}
   \langle V_2(\epsilon,p')|\,V_\mu\,|M_1(p)\rangle
   &=& {2 i\over m_1+m_2}\,\epsilon_{\mu\nu\alpha\beta}\,
    \epsilon^{*\nu} p'\mbox{}^\alpha p^\beta\,V(q^2) \,, \nonumber\\
   \langle V_2(\epsilon,p')|\,A_\mu\,|M_1(p)\rangle
   &=& \bigg[ (m_1+m_2)\,\epsilon_\mu^*\,A_1(q^2)
    - {\epsilon^*\cdot q\over m_1+m_2}\,(p+p')_\mu\,A_2(q^2)
    \nonumber\\
   &&\mbox{}- 2 m_2\,{\epsilon^*\cdot q\over q^2}\,q_\mu\,A_3(q^2)
    \bigg] + 2 m_2\,{\epsilon^*\cdot q\over q^2}\,q_\mu\,A_0(q^2)
    \,, \nonumber\\
\end{eqnarray}
where $\epsilon_\mu$ is the polarization vector, satisfying
$\epsilon\cdot p'=0$. Here, the form factor $A_3(q^2)$ is given by
the linear combination
\begin{equation}
   A_3(q^2) = {m_1+m_2\over 2 m_2}\,A_1(q^2)
   - {m_1-m_2\over 2 m_2}\,A_2(q^2) \,.
\end{equation}
Moreover, in order for the poles at $q^2=0$ to cancel, we must impose
the conditions
\begin{equation}
   F_1(0) = F_0(0) \,,\qquad A_3(0) = A_0(0) \,.
\end{equation}
The helicity amplitudes $H_0(q^2)$ and $H_{\pm}(q^2)$ are given by the
following combinations of form factors~\cite{Ba/Wi} ($K$ denotes the
momentum of the daughter meson in the parent rest frame):
\begin{eqnarray}
   H_\pm(q^2) &=& (m_1+m_2)\,A_1(q^2) \mp
    {m_1 K\over m_1+m_2}\,V(q^2) \,, \nonumber\\
   H_0(q^2) &=& {1\over 2 m_1\sqrt{q^2}}\,\bigg[
    (m_1^2-m_2^2-q^2) (m_1 + m_2)\,A_1(q^2) \nonumber\\
   &&\phantom{ {1\over 2 m_1\sqrt{q^2}}\,\bigg[ }
    \mbox{}- {4 m_1^2 K^2\over m_1+m_2}\,A_2(q^2) \bigg] \,.
\end{eqnarray}


\section*{References}

\begin{thebibliography}{999}


\bibitem {ba85}
M. Wirbel, B. Stech and M. Bauer, Z.\ Phys.\ C {\bf 29}, 637 (1985).

\bibitem {ISGW}
B. Grinstein, M.B. Wise and N. Isgur, Phys.\ Rev.\ Lett.\ {\bf 56},
298 (1986);\\
N. Isgur, D. Scora, B. Grinstein and M.B. Wise, Phys.\ Rev.\ D {\bf
39}, 799 (1989).

\bibitem {ISGW2}
D. Scora and N. Isgur, Phys.\ Rev.\ D {\bf 52}, 2783 (1995).

\bibitem {AlWo}
T. Altomari and L. Wolfenstein, Phys.\ Rev.\ Lett.\ {\bf 58}, 1563
(1987).

\bibitem {ko88}
J.G. K\"orner and G.A. Schuler, Z.\ Phys.\ C {\bf 38}, 511 (1988) [E:
{\bf 41}, 690 (1989)]; {\bf 46}, 93 (1990).

\bibitem {Jaus}
W. Jaus, Phys.\ Rev.\ D {\bf 41}, 142 (1990).

\bibitem {Casa}
R. Casalbuoni et al., Phys.\ Lett.\ B {\bf 292}, 371 (1992); {\bf 299},
139 (1993).

\bibitem {Alek}
R. Aleksan et al., Phys.\ Rev.\ D {\bf 51}, 6235 (1995);\\
A. Le Yaouanc, L. Oliver, O. Pene and J.C. Raynal, Phys.\ Lett.\ B {\bf
365}, 319 (1996).

\bibitem {Stechnew}
B. Stech, Phys.\ Lett.\ B {\bf 354}, 447 (1995);
Nucl.\ Phys.\ Proc.\ Suppl.\ {\bf 50}, 45 (1996).

\bibitem {Meli}
D. Melikhov, Phys.\ Rev.\ D {\bf 53}, 2460 (1996).

\bibitem {Faus}
R.N. Faustov, V.O. Galkin and A. Yu. Mishurov, Phys.\ Rev.\ D {\bf 53},
6302 (1996).

\bibitem {bs85}
M. Bauer and B. Stech, Phys.\ Lett.\ B {\bf 152}, 380 (1985).

\bibitem {bsw87}
M. Bauer, B. Stech and M. Wirbel, Z.\ Phys.\ C {\bf 34}, 103 (1987).

\bibitem {Shu1}
E.V. Shuryak, Phys.\ Lett.\ B {\bf 93}, 134 (1980); Nucl.\ Phys.\ B
{\bf 198}, 83 (1982).

\bibitem {Pasc}
J.E. Paschalis and G.J. Gounaris, Nucl.\ Phys.\ B {\bf 222}, 473
(1983);\\
F.E. Close, G.J. Gounaris and J.E. Paschalis, Phys.\ Lett.\ B {\bf
149}, 209 (1984).

\bibitem {Nuss}
S. Nussinov and W. Wetzel, Phys.\ Rev.\ D {\bf 36}, 130 (1987).

\bibitem {Volo}
M.B. Voloshin and M.A. Shifman, Yad.\ Fiz.\ {\bf 45}, 463 (1987)
[Sov.\ J.\ Nucl.\ Phys.\ {\bf 45}, 292 (1987)];
{\bf 47}, 801 (1988) [{\bf 47}, 511 (1988)].

\bibitem {Isgu}
N. Isgur and M.B. Wise, Phys.\ Lett.\ B {\bf 232}, 113 (1989);
{\bf 237}, 527 (1990).

\bibitem {review}
M. Neubert, Phys.\ Rep.\ {\bf 245}, 259 (1994);
Int.\ J.\ Mod.\ Phys.\ A {\bf 11}, 4173 (1996).

\bibitem {EiFe}
E. Eichten and F. Feinberg, Phys.\ Rev.\ D {\bf 23}, 2724 (1981).

\bibitem {CasL}
W.E. Caswell and G.P. Lepage, Phys.\ Lett.\ B {\bf 167}, 437 (1986).

\bibitem {Eich}
E. Eichten, in: Field Theory on the Lattice, edited by A. Billoire et
al., Nucl.\ Phys.\ B (Proc.\ Suppl.) {\bf 4}, 170 (1988).

\bibitem {Thac}
G.P. Lepage and B.A. Thacker, {\it ibid.} {\bf 4}, 199 (1988).

\bibitem {PoWi}
H.D. Politzer and M.B. Wise, Phys.\ Lett.\ B {\bf 206}, 681 (1988);
{\bf 208}, 504 (1988).

\bibitem {EiH1}
E. Eichten and B. Hill, Phys.\ Lett.\ B {\bf 234}, 511 (1990); {\bf
243}, 427 (1990).

\bibitem {Grin}
B. Grinstein, Nucl.\ Phys.\ B {\bf 339}, 253 (1990).

\bibitem {Geor}
H. Georgi, Phys.\ Lett.\ B {\bf 240}, 447 (1990).

\bibitem {Falk}
A.F. Falk, H. Georgi, B. Grinstein and M.B. Wise, Nucl.\ Phys.\ B
{\bf 343}, 1 (1990).

\bibitem {FGL}
A.F. Falk, B. Grinstein and M.E. Luke, Nucl.\ Phys.\ B {\bf 357},
185 (1991).

\bibitem {Mann}
T. Mannel, W. Roberts and Z. Ryzak, Nucl.\ Phys.\ B {\bf 368}, 204
(1992).

\bibitem {vcb}
M. Neubert, Phys.\ Lett.\ B {\bf 264}, 455 (1991);
{\bf 338}, 84 (1994).

\bibitem {Chay}
J. Chay, H. Georgi and B. Grinstein, Phys.\ Lett.\ B {\bf 247}, 399
(1990).

\bibitem {Bigi}
I.I. Bigi, N.G. Uraltsev and A.I. Vainshtein, Phys.\ Lett.\ B {\bf
293}, 430 (1992) [E: {\bf 297}, 477 (1993)];\\
I.I. Bigi, M.A. Shifman, N.G. Uraltsev and A.I. Vainshtein, Phys.\
Rev.\ Lett.\ {\bf 71}, 496 (1993);\\
I.I. Bigi et al., in: Proceedings of the Annual Meeting of the
Division of Particles and Fields of the APS, Batavia, Illinois, 1992,
edited by C.~Albright et al.\ (World Scientific, Singapore,
1993), p.~610.

\bibitem {MaWe}
A.V. Manohar and M.B. Wise, Phys.\ Rev.\ D {\bf 49}, 1310 (1994).

\bibitem {Adam}
M. Luke and M.J. Savage, Phys.\ Lett.\ B {\bf 321}, 88 (1994);\\
A.F. Falk, M. Luke and M.J. Savage, Phys.\ Rev.\ D {\bf 49}, 3367
(1994).

\bibitem {Blok}
B. Blok, L. Koyrakh, M.A. Shifman and A.I. Vainshtein, Phys.\ Rev.\
D {\bf 49}, 3356 (1994) [E: {\bf 50}, 3572 (1994)].

\bibitem {Thom}
T. Mannel, Nucl.\ Phys.\ B {\bf 413}, 396 (1994).

\bibitem {incltau}
A.F. Falk, Z. Ligeti, M. Neubert and Y. Nir, Phys.\ Lett.\ B {\bf
326}, 145 (1994).

\bibitem {shape}
M. Neubert, Phys.\ Rev.\ D {\bf 49}, 3392 and 4623 (1994).

\bibitem {Fermi}
I.I. Bigi, M.A. Shifman, N.G. Uraltsev and A.I. Vainshtein, Int.\
J.\ Mod.\ Phys.\ A {\bf 9}, 2467 (1994).

\bibitem {liferef}
I.I. Bigi et al., in: B Decays, edited by S. Stone, Second
Edition (World Scientific, Singapore, 1994), p.~134.

\bibitem {Gube}
B. Guberina, S. Nussinov, R. Peccei and R. R\"uckl, Phys.\ Lett.\ B
{\bf 89}, 111 (1979).

\bibitem {Bili}
N. Bilic, B. Guberina and J. Trampetic, Nucl.\ Phys.\ B {\bf 248},
261 (1984);\\
B. Guberina, R. R\"uckl and J. Trampetic, Z.\ Phys.\ C {\bf 33},
297 (1986).

\bibitem {ShiV}
M. Shifman and M. Voloshin, Sov.\ J.\ Nucl.\ Phys.\ {\bf 41}, 120
(1985); JETP {\bf 64}, 698 (1986).

\bibitem {Chern}
V. Chernyak, preprint BUDKERINP-94-69 (1994) [hep-ph/9407353].

\bibitem {Chris}
M. Neubert and C.T. Sachrajda, Nucl.\ Phys.\ B {\bf 483}, 339 (1997);\\
M. Neubert, in: Proceedings of 10th Les Rencontres de Physique de la
Vall\'ee d'Aoste: Results and Perspectives in Particle Physics, La
Thuile, Aosta Valley, March 1996, edited by M.~Greco (INFN Frascati
Physics Series, 1996), p.~245.

\bibitem {Beijing}
M. Neubert, in: Proceedings of the 17th International Symposium on
Lepton-Photon Interactions, Beijing, P.R. China, August 1995, edited
by Z. Zhi-Peng and C. He-Sheng (World Scientific, Singapore, 1996),
p.~298.

\bibitem {DeltaI}
W.A. Bardeen, A.J. Buras and J.-M. G\'erard, Phys.\ Lett.\ B {\bf
192}, 138 (1987); Nucl.\ Phys.\ B {\bf 293},787 (1987);\\
A.J. Buras, in: Non-perturbative Aspects of the Standard Model,
edited by J.~Abad et al., Nucl.\ Phys.\ B (Proc.\ Suppl.) {\bf 10A},
199 (1989).

\bibitem {Pichetal}
A. Pich and E. de Rafael, Phys.\ Lett.\ B {\bf 158}, 477 (1985);\\
A. Pich, B. Guberina and E. de Rafael, Nucl.\ Phys.\ B {\bf 277}, 197
(1986);\\
A. Pich, in: Hadronic Matrix Elements and Weak Decays, edited by
A.J.~Buras et al., Nucl.\ Phys.\ B (Proc.\ Suppl.) {\bf 7A}, 194
(1989);\\
A. Pich and E. de Rafael, Nucl.\ Phys.\ B {\bf 358}, 311 (1991).

\bibitem {Ne/St91}
B. Stech, in: Hadronic Matrix Elements and Weak Decays, edited by
A.J.~Buras et al., Nucl.\ Phys.\ B (Proc.\ Suppl.) {\bf 7A}, 106
(1989);\\
M. Neubert and B. Stech, Phys.\ Lett.\ B {\bf 231}, 477 (1989);
Ann.\ N.Y.\ Acad.\ Sci.\ {\bf 578}, 388 (1989);\\
M. Neubert, in: Proceedings of the International Symposium on Heavy
Quark Physics, Ithaca, New York, 1989, edited by P.S.~Drell and
D.L.~Rubin, AIP Conf.\ Proc.\ {\bf 196}, 52 (1989).

\bibitem {diquarks}
M. Neubert and B. Stech, Phys.\ Rev.\ D {\bf 44}, 775 (1991);\\
M. Neubert, in: The Hadron Mass Spectrum, edited by E.~Klempt and
K.~Peters, Nucl.\ Phys.\ B (Proc.\ Suppl.) {\bf 21}, 351 (1991);\\
B. Stech, Mod.\ Phys.\ Lett.\ A {\bf 6}, 3113 (1991);\\
B. Stech and Q.P. Xu, Z. Phys.\ C {\bf 49}, 491 (1991);\\
M. Neubert, Z.\ Phys.\ C {\bf 50}, 243 (1991).

\bibitem {Bj89}
J.D. Bjorken, in: New Developments in High-Energy Physics, edited by
E.G.~Floratos and A.~Verganelakis, Nucl.\ Phys.\ B (Proc.\ Suppl.)
{\bf 11}, 325 (1989).

\bibitem {Desh}
N. Deshpande, M. Gronau and D. Sutherland, Phys.\ Lett.\ B {\bf 90},
431 (1980);
Nucl.\ Phys.\ B {\bf 183}, 367 (1981).

\bibitem {Kama}
A.N. Kamal, Phys.\ Rev.\ D {\bf 33}, 1344 (1986).

\bibitem {Chau}
L.L. Chau and H.Y. Cheng, Phys.\ Rev.\ D {\bf 36}, 137 (1987).

\bibitem {Szcz}
A. Szczepaniak, E. Henley and S.J. Brodsky, Phys.\ Lett.\ B {\bf
243}, 287 (1990).

\bibitem {firstedi}
M. Neubert, V. Rieckert, B. Stech and Q.P. Xu, in: Heavy Flavours,
First Edition, edited by A.J. Buras and M.~Lindner (World Scientific,
Singapore, 1992), p.~286.

\bibitem {Dean}
A. Deandrea, N. Di Bartolomeo, R. Gatto and G. Nardulli, Phys.\ Lett.\
B {\bf 318}, 549 (1993).

\bibitem {Chen}
H.Y. Cheng, Phys.\ Lett.\ B {\bf 335}, 428 (1994).

\bibitem {Soar}
J.M. Soares, Phys.\ Rev.\ D {\bf 51}, 3518 (1995).

\bibitem {Orsa}
A. Le Yaouanc, L. Oliver, O. Pene and J.C. Raynal, Phys.\ Rev.\ D {\bf
52}, 2813 (1995).

\bibitem {Eber}
D. Ebert, R.N. Faustov and V.O. Galkin, preprint HUB-EP-96-67 (1997)
[hep-ph/9701218].

\bibitem {Rieck}
M. Neubert and V. Rieckert, Nucl.\ Phys.\ B {\bf 382}, 97 (1992).


\bibitem {wi69}
K.G. Wilson, Phys.\ Rev.\ {\bf 179}, 1499 (1969).

\bibitem {gi79}
F.G. Gilman and M.B. Wise, Phys.\ Rev.\ D {\bf 20}, 2392 (1979).

\bibitem {Sh77}
M.A. Shifman, A.I. Vainshtein and V.I. Zakharov, Nucl.\ Phys.\ B
{\bf 120}, 316 (1977).

\bibitem {Al74}
G. Altarelli and L. Maiani, Phys.\ Lett.\ B {\bf 52}, 351 (1974).

\bibitem {GaLe}
M.K. Gaillard and B.W. Lee, Phys.\ Rev.\ Lett.\ {\bf 33}, 108 (1974).

\bibitem {ACMP}
G. Altarelli, G. Curci, G. Martinelli and S. Petrarca, Phys.\ Lett.\
B {\bf 99}, 141 (1981); Nucl.\ Phys.\ B {\bf 187}, 461 (1981).

\bibitem {BuWe}
A.J. Buras and P.H. Weisz, Nucl.\ Phys.\ B {\bf 333}, 66 (1990).

\bibitem {G/W}
B. Grinstein, W. Kilian, T. Mannel and M.B. Wise, Nucl.\ Phys.\ B
{\bf 363}, 19 (1991).

\bibitem {E/mu}
M.J. Dugan and B. Grinstein, Phys.\ Lett.\ B {\bf 255}, 583 (1991).


\bibitem {fact1}
R.P. Feynman, in: Symmetries in Particle Physics, edited by A.
Zichichi, Acad.\ Press 1965, p.~167.

\bibitem {fact2}
O. Haan and B. Stech, Nucl.\ Phys.\ B {\bf 22}, 448 (1970).

\bibitem {wA1}
D. Fakirov and B. Stech, Nucl.\ Phys.\ B {\bf 133}, 315 (1978).

\bibitem {wA2}
L.L. Chau, Phys.\ Rep.\ {\bf 95}, 1 (1983).


\bibitem {Witt}
E. Witten, Nucl.\ Phys.\ B {\bf 160}, 57 (1979).

\bibitem {BGR}
A.J. Buras, J.M. G\'erard and R. R\"uckl, Nucl.\ Phys.\ B {\bf 268},
16 (1986).

\bibitem {BlokS}
B.Yu. Blok and M.A. Shifman, Yad.\ Fiz.\ {\bf 45}, 221, 478 and 841
(1987) [Sov.\ J.\ Nucl.\ Phys.\ {\bf 45}, 135, 301 and 522 (1987)];
{\bf 46}, 1310 (1987) [{\bf 46}, 767 (1987).

\bibitem {BlSh}
B. Blok and M. Shifman, Nucl.\ Phys.\ B {\bf 389}, 534 (1993); {\bf
399}, 441 and 459 (1993).

\bibitem {honreview}
T.E. Browder and K. Honscheid, Prog.\ Nucl.\ Part.\ Phys.\ {\bf 35}, 81
(1995).

\bibitem {Lewis}
J. Lewis, to appear in: Proceedings of the 2nd International Conference
on $B$ Physics and CP Violation, Honolulu, Hawaii, March 1997.

\bibitem {Kamal}
M. Gourdin, A.N. Kamal and X.Y. Pham, Phys.\ Rev.\ Lett.\ {\bf 73},
3355 (1994);\\
A.N. Kamal and F.M. Al-Shamali, preprint ALBERTA-THY-12-96 (1996)
[hep-ph/9605293].


\bibitem {Watson}
K.N. Watson, Phys.\ Rev.\ {\bf 88}, 1163 (1952).

\bibitem {Ka86}
A.N. Kamal, J. Phys.\ G (Nucl.\ Phys.) {\bf 12}, L43 (1986).

\bibitem {Rodri}
J.L Rodriguez, to appear in: Proceedings of the 2nd International
Conference on $B$ Physics and CP Violation, Honolulu, Hawaii, March
1997.

\bibitem {Rich}
J.D. Richman, preprint UCSB-HEP-97-01, to appear in: Proceedings of the
28th International Conference on High-Energy Physics (ICHEP 96),
Warsaw, Poland, July 1996 [hep-ex/9701014].


\bibitem {PDG}
R.M. Barnett et al., Review of Particle Properties, Phys.\ Rev.\ D
{\bf 54}, 1 (1996).

\bibitem {fDlimit}
J. Adler et al.\ (MARK III Collaboration), Phys.\ Rev.\ Lett.\ {\bf
60}, 1375 (1988).

\bibitem {aoki}
S. Aoki et al.\ (WA75 Collaboration), Prog.\ Theor.\ Phys.\ {\bf 89},
131 (1993).

\bibitem {franz}
D. Acosta et al.\ (CLEO Collaboration), Phys.\ Rev.\ D {\bf 49}, 5690
(1994);
D.~Gibaut et al.\ (CLEO Collaboration), conference paper
CLEO-CONF~95-22 (1995).

\bibitem {fDs3}
J.Z. Bai et al.\ (BES Collaboration), Phys.\ Rev.\ Lett.\ {\bf 74},
4599 (1995).

\bibitem {E653}
K. Kodama et al.\ (E653 Collaboration), Phys.\ Lett.\ B {\bf 382}, 299
(1996).

\bibitem {Evans}
H. Evans, to appear in: Tau 96, Proceedings of the 4th Workshop on Tau
Lepton Physics, Estes Park, Colorado, September 1996.

\bibitem {feta1}
H. Aihara et al.\ (TPC/$2\gamma$ Collaboration), Phys.\ Rev.\ Lett.\
{\bf 64}, 172 (1990).

\bibitem {feta2}
H.-J. Behrend et al.\ (CELLO Collaboration), Z.\ Phys.\ C {\bf 49},
401 (1991).


\bibitem {Braun}
A. Ali, V.M. Braun and H. Simma, Z.\ Phys.\ C {\bf 63}, 437 (1994).

\bibitem {BaBraun}
P. Ball and V.M. Braun, Phys.\ Rev.\ D {\bf 55}, 5561 (1997).

\bibitem {Bpirho}
CLEO Collaboration (J.P. Alexander et al.), Phys.\ Rev.\ Lett.\ {\bf
77}, 5000 (1996).

\bibitem {us}
G. Burdman, Z. Ligeti, M. Neubert and Y. Nir, Phys.\ Rev.\ D {\bf 49},
2331 (1994).

\bibitem {Kambor}
G. Burdman and J. Kambor, Phys.\ Rev.\ D {\bf 55}, 2817 (1997).


\bibitem {newrev}
T.E. Browder, K. Honscheid and D. Pedrini, Ann.\ Rev.\ Nucl.\ Part.\
Sci.\ {\bf 46}, 395 (1996).


\bibitem {bo90}
D. Bortoletto and S. Stone, Phys.\ Rev.\ Lett.\ {\bf 65}, 2951
(1990).

\bibitem {rosner}
J.L. Rosner, Phys.\ Rev.\ D {\bf 42}, 3732 (1990);

\bibitem {rie3}
V. Rieckert, Phys.\ Rev.\ D {\bf 47}, 3053 (1993).

\bibitem {qcd}
M. Neubert, Nucl.\ Phys.\ B {\bf 371}, 149 (1992); Phys.\ Rev.\ D
{\bf 46}, 2212 (1992).

\bibitem {sum92}
M. Neubert, Phys.\ Rev.\ D {\bf 46}, 3914 (1992).

\bibitem {Volker}
V. Rieckert, Ph.D.\ thesis, University of Heidelberg (1994),
unpublished.

\bibitem {matthias}
M. Neubert, Phys.\ Rev.\ D {\bf 46}, 1076 (1992).

\bibitem {abada}
A. Abada et al., Nucl.\ Phys.\ B {\bf 376}, 172 (1992).


\bibitem {St87}
B. Stech, Phys.\ Rev.\ D {\bf 36}, 975 (1987).

\bibitem {Ya89}
H.G. Dosch, M. Jamin and B. Stech, Z. Phys.\ C {\bf 42}, 167 (1989);\\
M. Jamin and M. Neubert, Phys.\ Lett.\ B {\bf 238}, 387 (1990).

\bibitem {iso}
B. Stech, in: Heavy Flavours, Nucl.\ Phys.\ B (Proc.\ Suppl.) {\bf 1B},
17 (1988).

\bibitem {Ball}
P. Ball and H.G. Dosch, Z.\ Phys.\ C {\bf 51}, 445 (1991).


\bibitem {Ba/Wi}
M. Bauer and M. Wirbel, Z. Phys.\ C {\bf 42}, 671 (1989).

\end{thebibliography}
\end{document}